\documentclass{article}
\usepackage{amsmath,amsthm,amssymb}

 \newtheorem{thm}{Theorem}[section]
 \newtheorem{defn}[thm]{Definition}
 \newtheorem{prop}[thm]{Proposition} 
 \newtheorem{rem}[thm]{Remark}
 \newtheorem{lem}[thm]{Lemma} 

\makeatother

\newcommand{\R}{\mathbb{R}}

\newcommand{\Z}{\mathbb{Z}}

\newcommand{\mst}{\R^{d+1}}

\newcommand{\cs}{\mathbf{\Sigma}}

\newcommand{\tpr}{\otimes\R^{2}}

\newcommand{\supp}{\textrm{supp}}

\newcommand{\subsp}[1]{\textrm{subsp}\left(#1\right)}

\newcommand{\idop}{\mathbf{1}}

\newcommand{\kleingordon}[1]{\left(\square+m^{2}\right)}

\newcommand{\hofb}[1]{\mathbf{H}\left(\mathbf{#1}\right)}

\begin{document}
\title{Relative Haag Duality for the Free Field\\ in Fock Representation}
\author{
Paolo Camassa\footnote{Supported by the EU network
"Quantum Spaces - Noncommutative Geometry"
HPRN-CT-2002-00280}\\
\small{Institut f\"ur Theoretische Physik, Georg-August Universit\"at G\"ottingen} \\
\small{Friedrich-Hund-Platz, 1 - 37077 - G\"ottingen - Germany}\\
\small{paolo.camassa@istruzione.it}
}
\maketitle

\begin{abstract}
We consider a natural generalization of Haag duality to the case
in which the observable algebra is restricted to a subset of the
space-time and is not irreducible: the commutant and the causal
complement have to be considered relatively to the ambient space.
We prove this relative form of Haag duality under quite general
conditions for the free scalar and electromagnetic field of space
dimension $d\geq 2$ in the vacuum representation. Such property is
interesting in view of a theory of superselection sectors for the
electromagnetic field.
\end{abstract}

\section{Introduction and Motivations}

In the framework of algebraic quantum field theory, the property
of Haag duality \cite{LQP}, which is a maximality condition with
respect to the locality principle, has proven to be a powerful
tool especially for the theory of superselection sectors. It has
been exploited in the DHR theory \cite{DHR1,DHR2,DHR3,DHR4}, where
superselection sectors are classified as unitary equivalence
classes of representations of the observable algebra satisfying a
selection criterion. In the presence of long range interactions,
like the electromagnetic field, the DHR criterion does not apply
and the problem of how to distinguish the physical charges in the
continuous infinity of inequivalent representations has to be
solved. According to a proposal of Buchholz \cite{BucSofQED}, the
electric charge should be recognizable when one looks at the
restriction of the representation of the observable algebra to a
forward light cone and considers unitary equivalence classes of
the representations restricted to any such cone.

From this point of view it is interesting to prove that the
naturally generalized version of Haag duality holds when the
ambient space-time is the forward light cone $V_+$. As a step in
this direction, we consider here the free scalar and
electromagnetic field on Minkowski space-time $\R^{d+1}$ in the
representation of the vacuum state, with the corresponding net of
von Neumann algebras $O\mapsto\mathfrak{R}\left(O\right)$, and
restrict our attention to a globally hyperbolic subset $M\subset\R
^{d+1}$ (as a special case $M=V_{+}$). Haag duality for double
cone regions in $\R^{d+1}$ is well known
\cite{AraFF1,AraFF2,DelSofAofFS,OstDforFBF,EckOstDforFBF,LeyRobTesDforQFF,%
HisLonMSofLAofFMSFT,HisDforLAinFQFT,BenNicLAforSFF&EMFF}. We prove
here its relative form, where the commutant is replaced by the
relative commutant with respect to the algebra
$\mathfrak{R}\left(M\right)$ and the relative causal complement of
a double cone is its causal complement inside $M$.

The technique used is essentially the following. From a general
theorem on Fock representation of the abstract algebra of
canonical commutation relations (CCR-algebra), recalled in Section
\ref{s:Fock-Representation}, we know that Haag duality is
equivalent to a duality property of the one-particle space. This
one-particle space duality is generally proven using a Cauchy
surface and the correspondence of solutions of the Klein-Gordon
equations with their Cauchy data. In Section 3, we consider a
globally hyperbolic space-time $M$, i.e. a space-time with a
Cauchy (smooth) surface $\cs$: e.g. $M$ can be $V_{+}$, a double
cone or a wedge region. We then consider the known structure of
the one-particle space associated to such a generic surface $\cs$.
In Section 4, for the scalar free field of any mass and
space-dimension $d\geq2$, we first prove relative duality for a
double cone $O$ and then for its relative commutant $O^{c}$,
Theorem \ref{thm:duality-scalar-field}. Notice that, unlike the
standard case, here duality for $O$ does not imply directly
duality for $O^{c}$. In Section 5, we consider the electromagnetic
free field and prove duality for a double cone $O$, Theorem \ref%
{thm:EM-duality-Oc}. In view of a theory of superselection sectors
for the electric charge, which can not be localized in a double
cone, one expects that a charged state can be seen in the vacuum
sector as the limit of states in which a charge is fixed and an
opposite charge is moved to space-like infinity. Staying inside
the forward light cone, the compensating charge can be moved along
a trajectory which, approaching the surface of the cone, goes to
light-like infinity with respect to the vertex of the cone and
space-like infinity with respect to any internal point of the
cone. To be more explicit, the compensating charge can be moved on
a hyperboloid along a specific direction and the charged state
obtained by this limit would be localized in a causally complete
open region containing the trajectory. It is useful, therefore, to
prove relative Haag duality for such a class of regions; this is
the content of Theorem \ref{thm:EM-rel-duality-A}.

\section{Fock Representation\label{s:Fock-Representation}}

\subsection{Net of von Neumann Algebras in Fock Representation}

The Weyl CCR-algebra $\mathfrak{A}\left(H,\sigma\right)$ is well defined as
an abstract C{*}-algebra canonically associated to the symplectic space $%
\left(H,\sigma\right)$ \cite{ManCCR,OAQSM2}. Whenever we have a
representation, or a state $\omega$ and the corresponding GNS representation
$\pi_{\omega}$, we can also define von Neumann algebras by closing $%
\mathfrak{A}\left(H,\sigma\right)$ and its C{*}-subalgebras with respect to
the weak topology.

When $H_{0}$ is a complex Hilbert space and the symplectic form $\sigma$ is
the imaginary part of the scalar product, we have the Fock representation of
$\mathfrak{A}\left(H_{0},\sigma\right)$ and, closing the local algebras $%
\mathfrak{A}\left(V\right)$, for any subspace $V$ of $H_{0}$, with respect
to the weak topology, we define a net of von Neumann algebras:
\begin{equation}
\subsp{H_{0}} \ni V\mapsto\mathfrak{R}\left(V\right):=%
\pi_{0}\left(\mathfrak{A}\left(V\right)\right)^{\prime\prime}\subset\mathcal{%
B}\left(\Gamma\left(H_{0}\right)\right),  \label{eq:vN-net-Fock}
\end{equation}
where $\subsp{H_{0}}$ indicates the set of \emph{closed}
real subspaces of $H_{0}$ and $\mathcal{B}\left(\Gamma\left(H_{0}\right)%
\right)$ is the algebra of bounded operators on the Fock space $%
\Gamma\left(H_{0}\right)$. We can actually define $\mathfrak{R}\left(V\right)
$ for any $V$, even not closed, but it coincides with the algebra associated
to its closure: $\mathfrak{R}\left(V\right)=\mathfrak{R}\left(\overline{V}%
\right)$ (see e.g. \cite{AraFF1} or \cite{LeyRobTesDforQFF}).

The set of closed subspaces is a complemented lattice, namely there is an
order relation $\subset$ by inclusion, with respect to which the operations
of supremum $\vee$ and infimum $\wedge$ are well defined, and an operation
of symplectic complement $^{\prime}$. The same is true for the set of von
Neumann algebras in $\mathcal{B}\left(\Gamma\left(H_{0}\right)\right)$,
where the complement is defined as the commutant. Notice that the complement
is a closed object in both cases, w.r.t. the respective topology.

The map in \ref{eq:vN-net-Fock} is a homomorphism of complemented lattices,
emphasized by the use of the same notations for the lattice operations in
both cases. This is the content of the following:

\begin{thm}
(Araki) \label{th:W*alg-in-Fock-space}Let $H_{0}$ be a complex
Hilbert space, $V,W,V_{\alpha}\in \subsp{H_{0}}$ (closed real
subspaces of $H_{0}$), then

\begin{enumerate}
\item $\mathfrak{R}\left(V\right)\supset\mathfrak{R}\left(W\right)$ $\iff$ $%
V\supset W$ and $\mathfrak{R}\left(V\right)=\mathfrak{R}\left(W\right)$ $\iff
$ $V=W$ (isotony),

\item $\vee_{\alpha}\mathfrak{R}\left(V_{\alpha}\right)=\mathfrak{R}%
\left(\vee_{\alpha}V_{\alpha}\right)$ (additivity),

\item \label{th:item:intersection}$\wedge_{\alpha}\mathfrak{R}%
\left(V_{\alpha}\right)=\mathfrak{R}\left(\wedge_{\alpha}V_{\alpha}\right)$,

\item \label{th:item:duality-irr}$\mathfrak{R}\left(V\right)^{\prime}=%
\mathfrak{R}\left(V^{\prime}\right)$ (abstract duality).
\end{enumerate}
\end{thm}

It was originally proven in \cite{AraFF1}, Th. 1; see also \cite%
{LeyRobTesDforQFF,AraYamQEofQFS}.

Property \ref{th:item:duality-irr} states that the commutant of the algebra
associated to a subspace is exactly the algebra associated to the symplectic
complement of that space. In \cite{LeyRobTesDforQFF} it is called \emph{%
abstract Haag duality}. We generalize now this concept of duality to the
context of a \emph{reducible} net of von Neumann algebras, when the
one-particle space is a \emph{real} subspace of a complex Hilbert space $%
H_{0}$.

\begin{defn}
Let $H$ be a real symplectic topological space and $\subsp{H} \ni
V\mapsto\mathfrak{R}\left(V\right)$ a net of von Neumann algebras.
For any $V\in \subsp{H}$, let
\[
V^{c}:=\left\{ f\in H\,:\,\sigma\left(f,g\right)=0 \quad \forall
g\in V\right\}
\]
be the symplectic complement in the ambient space $H$; for any
$\mathfrak{R} \left(V\right)$, let
\[
\mathfrak{R}\left(V\right)^{c}:=\left\{ A \in \mathfrak{R}
\left(H\right) \, : \, \left[A,B\right]=0 \quad \forall B \in
\mathfrak{R} \left(V\right)\right\}
\]
be the relative commutant in the ambient von Neumann algebra
$\mathfrak{R}\left(H\right)$.
\end{defn}

Note that the notations $V^{\prime}$ and $\mathfrak{R}\left(V\right)^{\prime}
$ were used to indicate the same complements in the special case in which $%
H=H_{0}$ is a complex Hilbert space and $\mathfrak{R}\left(H\right)$ is the
whole algebra of bounded operators on the representation space; thus the $%
\prime$-complement is an involutive operation, although the $c$-complement
is not in general.

\begin{defn}
\label{def:abstract-Haag-duality}A \emph{(not necessarily irreducible)} net
of represented von Neumann algebras $\subsp{H} \ni V\mapsto%
\mathfrak{R}\left(V\right)$ is said to satisfy abstract Haag
duality iff, for every $V\in \subsp{H}$,
\begin{equation} \nonumber
\mathfrak{R}\left(V\right)^{c}=\mathfrak{R}\left(V^{c}\right)
\end{equation}
\end{defn}

Let now choose a closed real subspace $H\subset H_{0}$ ($H$ is not
supposed to be invariant under the multiplication by $i$) such
that $\sigma$ is still non-degenerate when restricted to $H$; it
inherits the structure of a real Hilbert and symplectic space and
the net in eq. \ref{eq:vN-net-Fock} can be restricted to the net
of von Neumann algebras $\subsp{H} \ni
V\mapsto\mathfrak{R}\left(V\right)$ on the same representation
space, which
is in general no more irreducible. In that case one can write $%
V^{c}=V^{\prime}\cap H$ and $\mathfrak{R}\left(V\right)^{c}=\mathfrak{R}%
\left(V\right)^{\prime}\cap\mathfrak{R}\left(H\right)$.

\begin{prop} \label{pro:abstract-rel-haag-dual}
Let $H$ be a closed real subspace of a complex Hilbert space $H_{0}$, then
the net of von Neumann algebras $\subsp{H} \ni V\mapsto%
\mathfrak{R}\left(V\right)$ satisfies isotony, additivity and
abstract Haag
duality (Th. \ref{th:W*alg-in-Fock-space} and Def. \ref%
{def:abstract-Haag-duality}).
\end{prop}

\begin{proof}
Isotony and additivity are immediate consequences of the definition.
Concerning abstract Haag duality, one checks that
\begin{equation} \nonumber
\mathfrak{R}\left(V\right)^{c}:=\mathfrak{R}\left(V\right)^{\prime}\cap%
\mathfrak{R}\left(H\right)=\mathfrak{R}\left(V^{\prime}\right)\cap\mathfrak{R%
}\left(H\right)=\mathfrak{R}\left(V^{\prime}\cap H\right)=\mathfrak{R}%
\left(V^{c}\right),
\end{equation}
applying properties \ref{th:item:intersection} and \ref{th:item:duality-irr}
of Th. \ref{th:W*alg-in-Fock-space}.
\end{proof}

\section{Geometric Structure of the KG Equation}
\label{s:Geometric-Structure}

Let $M\subset\R^{d+1}$ be an open globally hyperbolic submanifold
of Minkowski space-time, maximal in the sense that there is no bigger submanifold
 with the same Cauchy surfaces; examples are the forward light cone $V_{+}$, a double cone or $\R%
^{d+1}$ itself. Global hyperbolicity is used to define a natural symplectic
structure on $C_{c}^{\infty}\left(M\right)$, corresponding to the commutator
of free fields, whilst translations in Minkowski space-time are used to
define an inner product, corresponding to the vacuum state (unique
translation invariant Fock state).

\subsection{Causal Structure in Minkowski Space-Time}

Minkowski space time has a natural causal structure:

\begin{defn}
\label{def:causal-compl} For every subset $O\subset\R^{d+1}$ we
define its causal complement as
\[
O^{\prime}:=\left\{ x\in \R^{d+1}:\left(x-y%
\right)^{2}<0\quad\forall y\in O\right\}
\]
where $x^2<0$ iff $x$ is spacelike. The causal completion of $O$
is $O^{\prime\prime}$.
\end{defn}

As global hyperbolicity has a crucial role in the following, we
will consider the family of regions that are causal completions of
open subsets of a Cauchy surface for $M$. Let $\cs$ be a Cauchy
surface, $\mathbf{B}\subset\cs$, the diamond
$C\left(\mathbf{B}\right)$ with base $\mathbf{B}$, defined as the
set of points for which any non space-like straight line
intersects $\mathbf{B}$, coincides with the causal completion of
$\mathbf{B}$:
$C\left(\mathbf{B}\right)=\mathbf{B}^{\prime\prime}$.
$C\left(\mathbf{B}\right)$ is the biggest region of space-time
whose observables are completely determined by those of
$\mathbf{B}$; the maximality of $M$ assumed above ensures that
$C\left(\mathbf{B}\right) \subset M$. In order to define spaces of
smooth functions on these regions we will consider open sets: it
can be easily proven that $\textrm{int}
\left(C\left(\mathbf{B}\right) \right)=
C\left(\mathbf{\textrm{int}\left(B\right)}\right)$ and therefore
$C\left(\mathbf{B}\right)$ is open when $\mathbf{B}$ is.

\begin{defn}
For every open subset $O\subset M $ we define $O^{c}$ as the
\emph{interior} of its (relative) causal complement in the ambient
manifold $M$, i.e.
\[
O^{c}=\textrm{int}\left\{ x\in
M:\left(x-y\right)^{2}<0\quad\forall y\in O\right\} \, ;
\]
clearly, $%
O^{c}= \textrm{int} \left(O^{\prime} \cap M \right)$.
\end{defn}

It follows easily form the definition that, using a Cauchy
surface, the causal complement is essentially reduced to set
complement: for every open subset $\mathbf{B}\subset\cs$, $C\left(\mathbf{B}\right)^{c}\cap\cs=\mathbf{%
\cs}\backslash\overline{\mathbf{B}}$ and hence
\begin{equation} \label{eq:rel-compl-set-compl}
C\left(\mathbf{B}\right)^{c}=C\left(\cs\backslash\overline{\mathbf{B}}\right)
\end{equation}

\subsection{Symplectic Space on Globally Hyperbolic Space-Times}

Following \cite{DimLO}, as $M$ is a globally hyperbolic $d+1$-dimensional
time-oriented space-time, there exist two unique continuous operators $%
E_{+},E_{-}$ (whose integral kernels are the advanced and retarded Green
functions) satisfying the following properties:
\begin{gather}
E_{\pm}:C_{c}^{\infty}\left(M\right)\rightarrow
C^{\infty}\left(M\right),
\notag \\
\left(\square+m^{2}\right)E_{\pm}=E_{\pm}\left(\square+m^{2}\right)=%
\mathbf{1}_{C_{c}^{\infty}\left( M \right)},  \label{eq:Green-functions} \\
\supp(E_{\pm}f)\subset J_{\pm}(\supp f),\notag
\end{gather}
where $J_{\pm}\left(O\right)$ is the set of all points that can be reached
from a point of the set $O$ with a forward/backward time-like or light-like
line; on Minkowski space-time their Fourier transforms are $\widehat{E_{\pm}}%
\left(p\right)=\frac{1}{p^{2}-m^{2}\pm ip_{0}\epsilon}$. We can then define
the propagator
\begin{equation}
E:=E_{+}-E_{-}:C_{c}^{\infty}(M)\rightarrow C^{\infty}(M),  \label{eq:E-def}
\end{equation}
which satisfies the Klein-Gordon equation and whose support is in the
time-like cone: $\supp(Ef)\subset J_{+}(\supp f)\cup J_{-}(%
\supp f)$. Its Fourier transform is $\widehat{E}\left(p\right)=-i%
\epsilon\left(p_{0}\right)\delta\left(p^{2}-m^{2}\right)$, where
$\epsilon\left(p_{0}\right)$ is the signum of $p_{0}$. The
transposes
are $E_{\pm}^{\prime}=E_{\mp}:C_{c}^{-\infty}\left(M\right)\rightarrow C^{-%
\infty}\left(M\right)$, indicating with the same symbol the extensions to
the dual spaces, and consequently $E^{\prime}=-E$.

For any choice of a Cauchy surface $\cs$, we have the following
continuous maps which associate to a solution of the Klein-Gordon
equation its Cauchy data:
\begin{equation}
\begin{array}{ccccc}
\rho_{0},\rho_{1} & : & C_{c}^{\infty}\left(M\right) & \rightarrow &
C_{c}^{\infty}\left(\cs\right) \\
&  & f & \mapsto & \rho_{0}f:=f|_{\cs},\,\rho_{1}f:=\frac{%
\partial}{\partial n}f|_{\cs} \, ,%
\end{array}
\label{eq:rho_0-rho_1}
\end{equation}
where $\frac{\partial}{\partial n}$ denotes the normal derivative
w.r.t. $\cs$.
The dual operators are $\rho_{0}^{\prime},\rho_{1}^{\prime}:C^{-\infty}\left(%
\cs\right)\rightarrow C^{-\infty}\left(M\right)$. We have the
following equality of maps from $C_{c}^{\infty}\left(M\right)$ to
$C^{\infty}\left(M\right)$ (see \cite{DimLO}):
\begin{equation}
E=E\left(\rho_{0}^{\prime}\rho_{1}-\rho_{1}^{\prime}\rho_{0}\right)E.
\label{eq:E-cauchy-data}
\end{equation}
Defining then, for $i,j\in \Z_{2}$, $E_{ij}:=\rho_{i+1}E\rho_{j+1}^{%
\prime}:C_{c}^{\infty}\left(\cs\right)\rightarrow C_{c}^{\infty}\left(%
\cs\right)$, we have
\begin{equation}
\begin{array}{cc}
E_{00}=E_{11}=\mathbf{0}, & E_{01}=-E_{10}=\mathbf{1}_{C_{c}^{\infty}\left(%
\cs\right)}%
\end{array}%
.  \label{eq:E-decomposition-cauchy-data}
\end{equation}

On the space $C_{c}^{\infty}\left(M\right)$, a natural bilinear
antisymmetric form is defined by
$\sigma\left(f,g\right):=\left\langle f,Eg\right\rangle $, where
$\left\langle \cdot,\cdot\right\rangle $ denotes the canonical
$L^{2}$ scalar product or, alternatively, the action of a function
as a distribution on another function. One can easily prove that
it is non-degenerate on the quotient space
$\frac{C_{c}^{\infty}\left(M\right)}{\sim}$, where $f\sim g$ iff
$Ef=Eg$. It defines therefore a symplectic form:
\begin{equation}
\begin{array}{ccccccc}
\sigma & : & \frac{C_{c}^{\infty}\left(M\right)}{\sim} & \times & \frac{%
C_{c}^{\infty}\left(M\right)}{\sim} & \rightarrow & \R \\
&  & \left[f\right] & , & \left[g\right] & \mapsto & \sigma\left(\left[f%
\right],\left[g\right]\right):=\left\langle f,Eg\right\rangle%
\end{array}%
.  \label{eq:sigma-global}
\end{equation}
The obvious isomorphic correspondence between equivalence classes
$\left[f\right]$ and solutions $Ef$ of the Klein-Gordon equation
can be pushed, given any Cauchy surface $\cs$, to the Cauchy data.
To every $Ef$ are uniquely associated its Cauchy data
$f_{0}=\rho_{0}Ef,\, f_{1}=\rho_{1}Ef$ on $\cs$. In the other
direction, for any couple of functions $%
f_{0},f_{1}\in C_{c}^{\infty}\left(\cs\right)$, there exists a function $%
f\in C_{c}^{\infty}\left(M\right)$ such that $\rho_{i}Ef=f_{i}$,
$i\in \Z_{2}$ (see again \cite{DimLO}). We have thus an
isomorphism
\begin{equation}
\frac{C_{c}^{\infty}\left(M\right)}{\sim}\simeq C_{c}^{\infty}\left(\cs%
\right) \oplus C_{c}^{\infty}\left(\cs\right) \,
\label{eq:isom-with-cauchy-data}
\end{equation}
and, in terms of the Cauchy data, using \ref{eq:E-cauchy-data} and
\ref{eq:E-decomposition-cauchy-data}, the symplectic form takes
the simplest form:
\begin{equation}
\sigma\left(f_{0}\oplus f_{1},g_{0}\oplus g_{1}\right)=\left\langle
f_{0},g_{1}\right\rangle -\left\langle f_{1},g_{0}\right\rangle .
\label{eq:sigma-cauchy-data}
\end{equation}

We can summarize the material in this subsection in the following

\begin{prop}
\label{pro:symplectic-spaces-M-S}To a time oriented globally hyperbolic
space-time $M$ is associated the symplectic space $\frac{%
C_{c}^{\infty}\left(M\right)}{\sim}$ of solutions of the Klein-Gordon
equation, for any mass $m$, with symplectic form \ref{eq:sigma-global}. To any Cauchy surface $%
\cs$ is associated the symplectic space of Cauchy data $%
C_{c}^{\infty}\left(\cs\right) \oplus
C_{c}^{\infty}\left(\cs\right)$
with symplectic form \ref{eq:sigma-cauchy-data}. They are isomorphic via \ref%
{eq:isom-with-cauchy-data}.
\end{prop}

\subsection{One-Particle Hilbert Space}

Let us now consider the embedding of $M$ in Minkowski space-time, where we
can use space-time translations and Fourier transform.

We can define in a natural way a complex structure on
$H_{0}:=H\left(\R^{d+1}\right)$, i.e. an operator $J$ which is a
symplectic square root of $\mathbf{-1}$ given by multiplication in Fourier transform by $%
i\epsilon\left(p_{0}\right)$: $\widehat{Jf}\left(p\right):=i\epsilon%
\left(p_{0}\right)\widehat{f}\left(p\right)$ (notice that multiplication by $%
i$ does not preserve the space of Fourier transforms of real functions). As $%
J$ is not a local operator, one cannot expect that $H\left(M\right)$ is
invariant under $J$, unless $H\left(M\right)=H_{0}$. The propagator $%
E:C_{c}^{\infty}(\R^{d+1})\rightarrow C^{\infty}(\R^{d+1})$ in
Minkowski space-time, when restricted to $M$, is exactly the
operator in eq. \ref{eq:E-def}, as it satisfies the required
properties. The operator
\begin{equation}
P=JE:C_{c}^{\infty}\left(\R^{d+1}\right)\rightarrow C^{\infty}\left(%
\R^{d+1}\right)  \label{eq:P-def}
\end{equation}
is positive as a quadratic form and is associated to the vacuum state $\omega
$ in the following way. On $C_{c}^{\infty}\left(\R^{d+1}\right)/\sim$%
, the bilinear form $\left(f,g\right)_{\R}:=\left\langle
f,Pg\right\rangle $ is a real scalar product and taking the
completion we have a complex Hilbert space
$H_{0}:=\overline{C_{c}^{\infty}\left(\R^{d+1}\right)/\sim}$, with
complex structure given by $J$. The symplectic form
$\sigma\left(f,g\right)=\left\langle f,Eg\right\rangle $ can be
extended to $H_{0}$ and there is a Fock state $\omega$ on the CCR-algebra $%
\mathfrak{A}\left(H_{0},\sigma\right)$ whose two point function is a complex
scalar product%
\begin{equation}
\omega\left(f,g\right)=\left(f,g\right)_{\R}+i\sigma\left(f,g\right)%
\,.  \label{eq:two-point-function}
\end{equation}

Let $\cs\subset M\subset\R^{d+1}$ be a Cauchy surface
for $M$ and $\rho_{i}$ the corresponding operators (cf. \ref{eq:rho_0-rho_1}%
). From the relation $P=JE$, restricted to $M$, and
\ref{eq:E-cauchy-data}, it follows that
\begin{eqnarray}
P=P\left(\rho_{0}^{\prime}\rho_{1}-\rho_{1}^{\prime}\rho_{0}\right)E & &
\text{on }C_{c}^{\infty}\left(M\right).  \label{eq:P-Cauchy-data}
\end{eqnarray}
Using this formula, one verifies that, in terms of the Cauchy data, $%
P$ has the form
\begin{equation}
\left\langle f,Pg\right\rangle =\left\langle f_{1},P_{11}g_{1}\right\rangle
+\left\langle f_{0},P_{00}g_{0}\right\rangle -\left\langle
f_{0},P_{01}g_{1}\right\rangle -\left\langle f_{1},P_{10}g_{0}\right\rangle ,
\label{eq:P-decomposition-cauchy-data}
\end{equation}
where, for $i,j\in \mathbb{Z}_{2}$, $P_{ij}:=\rho_{i+1}P\rho_{j+1}^{%
\prime}:C_{c}^{\infty}\left(\cs\right)\rightarrow C^{\infty}\left(\cs%
\right)$.

The space-time was assumed to be time-oriented. Time orientation enters in
the definition of $\sigma$ through the operator $E$: if we change the time
orientation, then $E\mapsto-E$, $\sigma\mapsto-\sigma$, the isomorphism in %
\ref{eq:isom-with-cauchy-data} is replaced by $\rho_{0}\mapsto\rho_{0}$, $%
\rho_{1}\mapsto-\rho_{1}$ and therefore
\begin{equation}
f_{0}\oplus f_{1}\mapsto f_{0}\oplus -f_{1}.
\label{eq:time-inversion-cauchy-data}
\end{equation}

Under time orientation inversion $J\mapsto -J$ and $P$ is invariant; it follows that \ref%
{eq:P-decomposition-cauchy-data} must be invariant under the map \ref%
{eq:time-inversion-cauchy-data} and thus
\begin{equation}
P_{10}=P_{01}=0.  \label{eq:P-off-diagonal-components}
\end{equation}
As a consequence, the real scalar product does not couple the 0-component
with the 1-component of the Cauchy data, which are real orthogonal:%
\begin{equation}
\left(f,g\right)_{\R}=\left\langle f,Pg\right\rangle =\left\langle
f_{0},P_{00}g_{0}\right\rangle +\left\langle
f_{1},P_{11}g_{1}\right\rangle .
\label{eq:real-scalar-product-cauchy-data}
\end{equation}

\begin{rem}
A time reversion map can be defined on $C_{c}^{\infty}\left(\R%
^{d+1}\right)/\sim$ using the map
\ref{eq:time-inversion-cauchy-data} on the right hand side of
\ref{eq:isom-with-cauchy-data}, but it depends on the choice of a
Cauchy surface, whilst the time orientation is a global
characteristic which enters in the definition of the symplectic
structure and does not depend on any $\cs$.
\end{rem}

We can summarize in the following

\begin{prop}
\label{pro:Hilbert-space-M-S}Let $M\subset\R^{d+1}$ be any
globally hyperbolic subset of Minkowski space-time, $\cs$ a Cauchy
surface for $M$, then the local real Hilbert space
$H\left(M\right)$, generated by functions with support in $M$, is
isomorphic to the space of Cauchy data:
\begin{equation}
H\left(M\right):=\overline{\frac{C_{c}^{\infty}\left(M\right)}{\sim}}%
^{\left\Vert \cdot\right\Vert }\simeq\overline{C_{c}^{\infty}\left(\cs%
\right)}^{\left\Vert \cdot\right\Vert _{0}}\oplus \overline{%
C_{c}^{\infty}\left(\cs\right)}^{\left\Vert \cdot\right\Vert
_{1}}, \label{eq:hilbert-space-cauchy-data}
\end{equation}
where the two norms are given by the real scalar products $%
\left(f_{0},g_{0}\right)_{0}:=\left\langle f_{0},P_{00}g_{0}\right\rangle $,
$\left(f_{1},g_{1}\right)_{1}:=\left\langle f_{1},P_{11}g_{1}\right\rangle $
and the direct sum is an orthogonal sum.
\end{prop}

\begin{rem}
Whilst $H_{0}$ is a complex Hilbert space and its symplectic form is the
imaginary part of the scalar product, the two isomorphic spaces in \ref%
{eq:hilbert-space-cauchy-data} are in general not invariant under $J$,
therefore they are only real symplectic Hilbert spaces, with the
corresponding symplectic forms given in Prop. \ref{pro:symplectic-spaces-M-S}%
.
\end{rem}

\section{Scalar Free Field}

\subsection{\label{ss:1-part-space-FreeField}Causal Structure of the
One-Particle Space}

The Hilbert space $H_{0}$ described in the previous section is the
one-particle space of the scalar free field on $d+1$-dimensional Minkowski
space-time. The quantum field and observable algebra are represented as
operators on the Fock space constructed on it (Section \ref%
{s:Fock-Representation}); $H:=H\left(M\right)\subset H_{0}$ is the
local
space associated to an open globally hyperbolic subset $M\subset\R%
^{d+1}$ of the $d+1$-dimensional Minkowski space-time.

As we have a natural association of CCR-subalgebras $\mathfrak{A}%
\left(V\right)$ to real subspaces $V\subset H$, the local structure of the
algebra of observables comes from the corresponding one on the one-particle
space $H$.

\begin{defn}
\label{def:1-p-Hilbert-space}For every open subset $O\subset M$ we define
the local subspace $H\left(O\right)\subset H$ as the real symplectic
(Hilbert) space $H\left(O\right):=\overline{C_{c}^{\infty}\left(O\right)/\sim%
}$. For every open subset $\mathbf{B}\subset\cs$ we define the
local subspace $\mathbf{H}\left(\mathbf{B}\right):=\overline{%
C_{c}^{\infty}\left(\mathbf{B}\right)}\oplus \overline{C_{c}^{\infty}\left(%
\mathbf{B}\right)}$.
\end{defn}

The local subspaces are not complex spaces because they are not invariant
under the multiplication by $i$. According to the isomorphisms of Prop. \ref%
{pro:symplectic-spaces-M-S} and \ref{pro:Hilbert-space-M-S}
\begin{equation}
H\left(C\left(\mathbf{B}\right)\right)=\overline{C_{c}^{\infty}\left(C\left(%
\mathbf{B}\right)\right)/\sim}=\bigcap_{A\supset\mathbf{B}%
}H\left(A\right)\simeq\overline{C_{c}^{\infty}\left(\mathbf{B}\right)\otimes%
\R^{2}}=\mathbf{H}\left(\mathbf{B}\right).
\label{eq:isom-cauchy-data-local}
\end{equation}
The second equality follows from the fact that for any open $A$ such that $%
\mathbf{B}\subset A\subset C\left(\mathbf{B}\right)$, obviously $%
H\left(A\right)\subset H\left(C\left(\mathbf{B}\right)\right)$ but also $%
H\left(A\right)\supset H\left(C\left(\mathbf{B}\right)\right)\simeq\mathbf{H}%
\left(\mathbf{B}\right)$ with the same argument used in the proof of Prop. \ref%
{pro:symplectic-spaces-M-S}. Using the Cauchy data formulation one
can exploit eq. \ref{eq:rel-compl-set-compl} and thus
\begin{equation}
H\left({C\left(\mathbf{B}\right)}^{c}\right)\simeq\mathbf{H}\left({\cs \backslash \overline{%
\mathbf{B}}}\right).  \label{eq:isom-cauchy-data-local2}
\end{equation}

\subsection{\label{ss:local-CCR-alg}Haag Duality and One-Particle-Space
Duality}

Having defined a local structure of the one-particle space
$H\subset H_{0}$, the CCR-algebra
$\mathfrak{A}\left(H,\sigma\right)$ gains a local structure.
If we then consider the Fock representation on the Hilbert space $%
\Gamma\left(H_{0}\right)$, we can define a local net of von Neumann algebras
$O\mapsto\mathfrak{R}\left(O\right)$.

\begin{defn}
For every open subset $O\subset M$, the local von Neumann algebra $\mathfrak{%
R}\left(O\right)$ in Fock representation is the algebra $\mathfrak{R}%
\left(H\left(O\right)\right)$ (eq. \ref{eq:vN-net-Fock}), i.e. the
algebra generated by the Weyl operators $W\left(f\right)$ for all
smooth $f$ with compact support in $O$.
\end{defn}

From the properties of Fock representation and of the one-particle space, we
easily deduce the properties of the net of local von Neumann algebras $\mathfrak{R}$%
. E.g., the algebra $\mathfrak{R}\left(C\left(\mathbf{B}\right)\right)$
associated to a diamond $C\left(\mathbf{B}\right)$ coincides with the
algebra $\mathfrak{R}\left(\mathbf{B}\right):=\mathfrak{R}\left(\mathbf{H}%
\left(\mathbf{B}\right)\right)$ associated to its basis: $\mathfrak{R}\left(%
\mathbf{B}\right)=\bigwedge_{A\supset\mathbf{B}}\mathfrak{R}\left(A\right)=%
\mathfrak{R}\left(C\left(\mathbf{B}\right)\right)$, where $A$
ranges over the set of open subsets of $\mst$.

Abstract Haag duality (Th. \ref{th:W*alg-in-Fock-space}) implies
that Haag duality, even in its relative form, is equivalent to a
property of the local one-particle Hilbert space that we call
one-particle space duality. It is again a maximality property with
respect to locality, namely the fact that the local space
$H\left({O}^{c}\right)$ associated to the causal complement
$O^{c}$ of a region $O$ of space-time coincides with the
symplectic complement $H\left(O\right)^{c}$ of the local space
associated to $O$.

\begin{defn}
The local subspaces $H\left(O\right)\text{ or }\mathbf{H}\left(\mathbf{B}%
\right)\subset H$ are said to satisfy one-particle space duality
iff, respectively,
\begin{equation*}
H\left({O}^{c}\right)^{c}=H\left(O\right)\,, \quad\quad\quad
\mathbf{H}\left({\cs\backslash\overline{\mathbf{B}}}\right)^{c} =
\mathbf{H}\left(\mathbf{B}\right)\,.
\end{equation*}
\end{defn}

\begin{rem}
\label{rem:duality-sympl-compl}The previous equations are not equivalent to
\begin{equation*}
H\left(O\right)^{c}=H\left({O}^{c}\right)\,,\quad\quad\quad\mathbf{H}\left(%
\mathbf{B}\right)^{c}=\mathbf{H}\left({\cs\backslash\overline{\mathbf{B}}}\right)\,.
\end{equation*}
as the symplectic complement is not in general an involutive map on closed
subspaces: $V^{cc}$ is in general not equal to $V$ even for closed $V\subset
H$.
\end{rem}

Such a property is expected to hold only for a certain class of
regions; indeed, some regularity of the region $O$ is needed to
prove the following property that will be used in the proof of Th.
\ref{thm:ops-duality}:

\begin{defn}
\label{def:outer-regularity}We say that a local subspace $\mathbf{H}\left(%
\mathbf{B}\right)$ or $H\left(O\right)$ satisfies outer regularity
if:
\begin{equation*}
\mathbf{H}\left(\mathbf{B}\right)=\bigcap_{\mathbf{A\supset\overline{B}}}%
\mathbf{H}\left(\mathbf{A}\right)\,,\quad H\left(O\right)=\bigcap_{A\mathbf{%
\supset}\overline{O}}H\left(A\right)
\end{equation*}
where $\mathbf{A}$ ranges over the open neighborhoods of $\overline{\mathbf{B%
}}$ in $\cs$ and $A$ over the open diamonds containing $%
\overline{O}$ in $M$.
\end{defn}

A similar definition can be given for local algebras. We quote the
main statement here and refer to App. \ref{app:Outer-Regularity}
for proofs and references.

\begin{prop}
\label{pro:outer-regularity}Let $\mathbf{B}\subset\cs$ be the
basis of a double cone $O\subset\R^{d+1}$, then the local spaces $%
\mathbf{H}\left(\mathbf{B}\right)\simeq H\left(O\right)$ satisfy
outer regularity.
\end{prop}

A simple Lemma leads us to the main theorem:

\begin{lem} \label{lem:duality-ops-duality}
Relative duality $\mathfrak{R}%
\left(O^{c}\right)^{c}=\mathfrak{R}\left(O\right)$ is equivalent
to one-particle space duality
$H\left({O}^{c}\right)^{c}=H\left(O\right)$ or,
in terms of Cauchy data, $\mathbf{H}\left({\cs \backslash \overline{\mathbf{B}}}%
\right)^{c}=\mathbf{H}\left(\mathbf{B}\right)$.
\end{lem}

\begin{proof}
From Prop. \ref{pro:abstract-rel-haag-dual},
$\mathfrak{R}\left(H\left(O^{c}\right)\right)^{c}=%
\mathfrak{R}\left(H\left({O}^{c}\right)^{c}\right)$; to use Cauchy
data, apply the isomorphisms in \ref{eq:isom-cauchy-data-local} and \ref%
{eq:isom-cauchy-data-local2}.
\end{proof}

We state now relative Haag duality in some relevant cases:

\begin{thm} \label{thm:duality-scalar-field}
Let the ambient space $M$ be any open globally hyperbolic
submanifold of Minkowski space-time $\R^{d+1}$ with $d\geq2$, $O$
any double cone with $\overline{O}\subset M$, then the algebra of
observables for the free scalar field in the representation of the
vacuum state satisfies relative duality:
\begin{equation*}
\mathfrak{R}\left(O^{c}\right)^{c}=\mathfrak{R}\left(O\right)
\end{equation*}

The dual relation (cf. Rem. \ref{rem:duality-sympl-compl})
\begin{equation*}
\mathfrak{R}\left(O\right)^{c}=\mathfrak{R}\left(O^{c}\right)
\end{equation*}
holds in the following cases

\begin{itemize}

\begin{item}
the field is massive
\end{item}

\begin{item}
the field is massless, $M \equiv V_{+}$ and
$O=C\left(\mathbf{B}\right)$ is any double cone with basis
$\mathbf{B}\subset \cs$, where $\cs=\left\{ x\in\R^{d+1} : x_0\geq
0, x^2=c\in \R_+ \right\}$ is an hyperboloid in $V_{+}$.
\end{item}

\end{itemize}
\end{thm}

\begin{proof}
The proof follows from Lemma \ref{lem:duality-ops-duality} and the
theorems below on one-particle space duality: Th.
\ref{thm:ops-duality} and, for the dual relation, considerations
in Subsection \ref{sss:massive-case} and Th.
\ref{thm:ops-duality2}.
\end{proof}

\subsection{\label{sub:Duality-for-Oc}Duality for a Double Cone $O$}

We want to prove that $H\left({O}^{c}\right)^{c}=H\left(O%
\right)$ for any double cone $O\subset\overline{O}%
\subset M\subset\R^{d+1}$. We use a Cauchy surface: let $O_{1}$ be a double cone such that $%
\overline{O}\subset O_{1}\subset M$, choose a Cauchy surface $\cs
$ containing the basis $\mathbf{B}_{1}$ of $O_{1}=C\left(\mathbf{B}_1%
\right)\subset\R^{d+1}$ (and the basis $\mathbf{B}$ of $O$). Such
a Cauchy surface can be explicitly constructed in the cases of
interest ($M=V_+
$) and its existence has been proved in full generality in \cite%
{SanBerEoCStSCS}. In terms of Cauchy data the previous equality is
equivalent to $\mathbf{H}\left({\cs \backslash \overline{\mathbf{B}}}^{c}\right)^{c}=%
\mathbf{H}\left(\mathbf{B}\right)$. Following \cite%
{LeyRobTesDforQFF,LudRobLQE&AVS}, we prove duality via outer regularity, by
proving that
\begin{equation}
\mathbf{H}\left({\cs \backslash \overline{\mathbf{B}}}\right)^{c}=\bigcap_{\mathbf{%
A\supset\overline{B}}}\mathbf{H}\left(\mathbf{A}\right)\,.
\label{eq:duality-outer-reg}
\end{equation}
Using the symplectic form $\sigma$, we can define in a natural way an
immersion
\begin{equation} \label{eq:phi-map-to-distrib}
\psi: H\supset C_{c}^{\infty}\left(\cs\right)\tpr \hookrightarrow%
C^{-\infty}\left(\cs\right)\tpr
\end{equation}
with $\left\langle
\psi\left(f\right),g\right\rangle:=\sigma\left(f,g\right)$ for any
$g \in C_{c}^{\infty}\left(\cs\right)\tpr$. This map on
$C_{c}^{\infty}\left(\cs\right)\tpr$ is
continuous, because $f_{n}\overset{\left\Vert \cdot\right\Vert }{\rightarrow}%
f$ implies $\sigma\left(f_{n},g\right)\rightarrow\sigma\left(f,g\right)$ for
any $g\in C_{c}^{\infty}\left(\cs\right)\tpr$, i.e. $%
\psi\left(f_{n}\right)\rightarrow\psi\left(f\right)$ in the
topology of distributions (weak convergence); it can therefore be
extended to the whole Hilbert space $H$. It is injective, because
if $\sigma\left(f,g\right)=0$ for any $g\in
C_{c}^{\infty}\left(\cs\right)\tpr$, which
is dense in $H$, then $f=0$. Restricted to $C_{c}^{\infty}\left(\cs%
\right) \tpr$, the map $\psi$ is just the composition of the map
$f_{0}\oplus f_{1} \overset{U}{\mapsto } -f_{1}\oplus f_{0}$ with
the canonical injection $C_{c}^{\infty}\left(\cs\right) \tpr
\hookrightarrow C^{-\infty}\left(\cs\right) \tpr$: whenever
$f,g\in C_{c}^{\infty}\left(\cs\right) \tpr$%
, according to \ref{eq:sigma-cauchy-data},%
\begin{equation}
\left\langle Uf,g\right\rangle =\left\langle -f_{1}\oplus
f_{0},g_{0}\oplus g_{1}\right\rangle =\left\langle
f_{0},g_{1}\right\rangle -\left\langle f_{1},g_{0}\right\rangle
=\sigma\left(f,g\right)=\left\langle
\psi\left(f\right),g\right\rangle \,.
\label{eq:phi-Fourier-antitransform}
\end{equation}

\begin{rem}
The former construction shows that a general element
$f=f_{0}\oplus f_{1}\in H$, as a limit of functions, is a
distribution $f_{0}\oplus f_{1}
\in \mathcal{S}^{\prime}\oplus \mathcal{S}^{\prime}$ and $%
\psi\left(f\right)=-f_{1}\oplus f_{0}$.
\end{rem}

\begin{thm}
\label{thm:ops-duality}For any double cone $O$ such that $\overline{O}%
\subset M$ with basis $\mathbf{B}$, the local subspaces $H\left(O\right)%
\simeq\mathbf{H}\left(\mathbf{B}\right)$ satisfy (relative) duality:
\begin{equation}
H\left({O}^{c}\right)^{c}=H\left(O\right)\,,\quad\quad\quad
\mathbf{H}\left(\cs\backslash\overline{\mathbf{B}}\right)^{c}=
\mathbf{H}\left(\mathbf{B}\right)\,.
\label{eq:one-particle-space-duality-Oc}
\end{equation}
\end{thm}

\begin{proof}
Let $f\in
\mathbf{H}\left(\cs\backslash\overline{\mathbf{B}}\right)^{c}$,
then, by definition of the relative symplectic complement of
$\mathbf{H}\left(\cs\backslash\overline{\mathbf{B}}\right)$, the
distribution $\psi\left(f\right)$ has support in
$\overline{\mathbf{B}}\subset\mathbf{B}_{1}$. Now, if we
restrict our attention to the region $O_{1}$, the Cauchy surface $\mathbf{%
\cs}$ can be identified with $\R^{d}$ and we can use the known
results on Minkowski space-time and time-$0$ Cauchy surface.

To find regular functions approximating $f\in\mathbf{H}\left(\cs \backslash
\overline{\mathbf{B}}\right)^{c}$ we use convolutions with regular functions $%
\rho_{n}$ approximating Dirac's $\delta$. Let $\rho\in C_{c}^{\infty}\left(%
\R^{d}\right)$ be an even function such that $\int\rho\left(\mathbf{x%
}\right)\, d\mathbf{x}=1$, and $C_{c}^{\infty}\left(\R%
^{d}\right)\ni\rho_{n}\left(x\right):=n^{d}\rho\left(nx\right)$, then define
$f_{n}:=\rho_{n}\ast\psi\left(f\right)$. The functions $f_{n}$ are in $%
C_{c}^{\infty}\left(\R^{d}\right)\tpr$ and have for large $n$
support in $\mathbf{B}_{1}$. Therefore, for $n$ big enough, as
elements of $C_{c}^{\infty}\left(\cs\right)\tpr$, they belong to
$\mathbf{H}\left(\mathbf{B_{1}}\right)\simeq H\left(O_{1}\right)$.
It can be proved, using Fourier transform (see
\cite{LeyRobTesDforQFF}) that $f_{n}\rightarrow f$ in
$\mathbf{H}\left(\mathbf{B_{1}}\right)$. Moreover,
for any open set $\mathbf{A}\ni0$, $f_{n}\in \mathbf{H}\left(\mathbf{B+A}%
\right)$ for large $n$, because $\supp f_{n}=\supp
\psi\left(f\right)+\supp \rho_{n}\subset\mathbf{B}+\mathbf{A}$.
Therefore, $f=\lim f_{n}$ is in $\bigcap_{\mathbf{A}}\mathbf{H}\left(\mathbf{%
B+A}\right)$, where $\mathbf{A}$ ranges in the set of open neighborhoods of $%
0$ in $\R^{d}$. Finally, by outer regularity, Prop. \ref%
{pro:outer-regularity}, we conclude that $f\in \mathbf{H}\left(\mathbf{B}%
\right)$.
\end{proof}

\subsection{Duality for $O^{c}$}

We address now the other duality relation $\mathfrak{R}\left(O\right)^{c}=%
\mathfrak{R}\left(O^{c}\right)$, where the role of $O$ and $O^{c}$ are
interchanged, recalling Rem. \ref{rem:duality-sympl-compl}. This relation is
equivalent to $H\left(O\right)^{c}=H\left({O}^{c}\right)$ (or $%
H\left(O\right)^{\prime}\cap H\left(M\right)=H\left(O^{\prime}\cap M\right)$%
). Difficulties arise because the elements of $H\left(O\right)^{c}$ have
support in $O^{c}$, which is not compact if $M$ is not, and possibly because
of the topology of $O^{c}$. We do not prove this relation in generality but
we specialize now to the case $M=V_{+}$ and distinguish the massive and
massless cases.

\subsubsection{The massive case: $H\left(V_{+}\right)=H_{0}$}
\label{sss:massive-case}

In the massive case, the algebra associated to the forward (or backward)
light-cone is already the total algebra of observables or, in terms of
one-particle spaces, $H\left(V_{+}\right)=H\left(\R%
^{d+1}\right)=H_{0}$ is the whole Hilbert space; see Theorems 1 and 2 in
\cite{SW-TSiQFT}.

In such case, the relative symplectic complement coincides with the
symplectic complement ($V^{c}=V^{\prime} \;\; \forall V\in \text{subsp}%
\left(H\left(V_{+}\right)\right)$) and is therefore involutive,
$V^{cc}=V$. Duality for $O^{c}$ follows trivially from duality for
$O$ because, taking
the dual of both sides in eq. \ref{eq:one-particle-space-duality-Oc}, where $%
\mathbf{H}\left({\cs \backslash \overline{\mathbf{B}}}\right)^{c}\equiv\mathbf{H}\left({%
\cs\backslash\overline{\mathbf{B}}}\right)^{\prime}$, we obtain $\mathbf{H}\left({%
\cs\backslash\overline{\mathbf{B}}}\right)=\mathbf{H}\left(\mathbf{B}\right)^{\prime}$.

\subsubsection{\label{sub:massless-case}The massless case}

In the massless case with odd space dimension $d$,
$H\left(V_{+}\right)$ has a big symplectic complement
$H\left(V_{+}\right)^{\prime}=H\left(V_{-}\right) $ in $H_{0}$.
This is a consequence of Huygens' principle and corresponds
to time-like duality for the algebra of the light-cone: $\mathfrak{R}%
\left(V_{+}\right)^{\prime}=\mathfrak{R}\left(V_{-}\right)$ (see \cite%
{BucLQFwNTI}).

However, in the massless case, for any space-time dimension $d+1$,
we can use conformal covariance and the consequent unitary
equivalence of the algebras $\mathfrak{R}\left(V_{+}\right)$ and
$\mathfrak{R}\left(O_{1}\right) $, where $O_{1}$ is a double cone
\cite{HisLonMSofLAofFMSFT}, thus reducing our ambient space to a
relatively compact one.

A relativistic ray inversion map $\varphi_{0}:x\mapsto-x/x^{2}$ transforms a
double cone $O_{1}$, which has the upper vertex in $0$ and the lower vertex
in $\left(-T,\mathbf{0}\right)$, to the translated forward light cone $%
V_{+}+\left(\frac{1}{T},\mathbf{0}\right)$ and its basis $\mathbf{B}%
_{1}=\left\{ \left(-\frac{T}{2},\mathbf{x}\right):\left\Vert \mathbf{x}%
\right\Vert <\frac{T}{2}\right\} $ to the hyperboloid $\left\{ x:x_0\geq
0,\, x^{2}=1/T^{2}\right\} +\left(\frac{1}{T},\mathbf{0}\right)$. Of course,
as $\varphi_{0}^{-1}=\varphi_{0}$, also the converse is true and composing
the inversion map $\varphi_{0}$ with suitable translations a conformal map $%
\varphi$ can be found such that $O_{1}:=\varphi\left(V_{+}\right)$ is a
double cone and, for any double cone $O$ with closure in $V_{+}$, $%
\varphi\left(O\right)$ is a double cone with closure in $O_{1}$. For further
simplicity, we assume that the vertices of $O$ and the vertex of $V_{+}$ lie
on the same straight line. In this case an hyperboloid $\cs%
=\left\{ x:x_0\geq 0, x^{2}=c\in \R_{+}\right\} $ can be found
such that $O=C\left(\mathbf{B}\right)$ with $\mathbf{B}\subset\cs$
and the map $\varphi$ can be chosen so that the basis of $O_{1}$, $\mathbf{B}%
_{1}=\varphi\left(\cs\right)$, is lying on the time-$0$ surface $%
\R^{d}\subset\R^{d+1}$. $\mathbf{B}_{1}\subset\R^{d}$ is obviously
a Cauchy surface for $O_{1}$, whilst $\R^{d}$ is the standard
time-$0$ Cauchy surface for $\R^{d+1}$.

\begin{thm} \label{thm:ops-duality2}
Let $\cs=\left\{ x\in\R^{d+1} :
x_0\geq 0, x^2=c\in \R_+ \right\}$ be an hyperboloid in $V_{+}$, $%
O=C\left(\mathbf{B}\right)$ any double cone with basis $\mathbf{B}\subset%
\cs$, then the local subspaces $H\left(O^{c}\right)\simeq\mathbf{%
H}\left(\mathbf{\cs\backslash\overline{B}}\right)$ satisfy
(relative) duality:
\begin{equation}
H\left({O}^{c}\right)=H\left(O\right)^{c}\,,\quad\quad\quad\mathbf{H}\left({%
\cs\backslash\overline{\mathbf{B}}}\right)=\mathbf{H}\left(\mathbf{B}\right)^{c}\,.
\label{eq:1-p-s-duality-Oc}
\end{equation}
\end{thm}

\begin{proof}
As explained above, using a conformal transformation, we can
reduce the
problem to the following: let $O=C\left(\mathbf{B}\right)$ and $O_{1}=C\left(%
\mathbf{B}_{1}\right)$ be two double cones with basis $\mathbf{B}$ and $%
\mathbf{B}_{1}$ such that $\overline{\mathbf{B}}\subset\mathbf{B}_{1}\subset%
\cs\equiv\R^{d}$, then we want to prove that $H\left(%
\mathbf{B}\right)^{\prime}\cap H\left(\mathbf{B}_{1}\right)=H\left(\mathbf{B}%
^{\prime}\cap\mathbf{B}_{1}\right)$. Let $f\in \mathbf{H}\left(\mathbf{B}%
\right)^{\prime}\cap\mathbf{H}\left(\mathbf{B_{1}}\right)$, applying the map %
\ref{eq:phi-map-to-distrib} we have a distribution
$\psi\left(f\right)\in C^{-\infty}\left(\R^{d}\right)\tpr$ whose
support
must be in $\overline{\mathbf{B}_{1}}$ because $f\in \mathbf{H}\left(\mathbf{%
B_{1}}\right)$ and in $\mathbf{B}^{c}$ because $f\in \mathbf{H}\left(\mathbf{%
B}\right)^{\prime}$, i.e. $\supp f\subset\overline{\mathbf{B}_{1}}%
\backslash\mathbf{B}$. Let $\rho,\rho_{n}\in C_{c}^{\infty}\left(\R%
^{d}\right)$ be as in Sec. \ref{sub:Duality-for-Oc}, $f_{n}:=\rho_{n}\ast%
\psi\left(f\right)$. The functions $f_{n}$ are in $C_{c}^{\infty}\left(%
\R^{d}\right)\tpr$ and, for any open neighborhood $%
\mathbf{A}\subset\R^{d}$ of $0$, have support in
$\left(\mathbf{B}_{1}\backslash\overline{\mathbf{B}}\right)+\mathbf{A}$,
therefore belong to
$\hofb{\left(B_{1}\backslash\overline{B}\right)+A}$. As in Sec.
\ref{sub:Duality-for-Oc}, $f_{n}\rightarrow f$ in $\hofb{\R^{d}}$,
hence $f=\lim f_{n}\in \bigcap_{\mathbf{A}}\mathbf{H}\left(\mathbf{%
\left(B_{1}\backslash\overline{B}\right)+A}\right)$, and one-particle %
space duality follows from outer regularity for the non
contractible region $B_{1}\backslash\overline{B}$,
 Prop. \ref{pro:outer-regularity-O1-Oc} in App. \ref{app:Outer-Regularity}.
\end{proof}

\section{Free Electromagnetic Field}

Let the ambient space $M\subset\R^{d+1}$ be an open globally
hyperbolic maximal submanifold, as in Sect.
\ref{s:Geometric-Structure}, with the further requirement that it
is \emph{contractible} to avoid cohomological problems. We denote
by $\Omega_{}^{\ast}\left(M\right)$ the set of differential forms,
with a prefix $Z^d,Z^\delta$ for the closed forms with respect to
$d$ or its formal adjoint $\delta$ and similarly $B^d,B^\delta$
for the exact forms (cf. App. \ref{app:Differential-forms} for
more details).

\subsection{Quantum Free Electromagnetic Field}

The quantum electromagnetic field $F$ is an operator valued
continuous functional on the test space of $2$-forms satisfying
Maxwell equations, i.e. $dF=0$ and $\delta F=0$. From the first
equation, as the $2$-cohomology of $M $ is trivial (the cohomology
of currents coincides with the cohomology of forms, generalized De
Rham's Theorem, Chap. IX in \cite{SchwartzTdD}), it is the
external derivative of the (equivalence class $\left[A\right]$ of
the) electromagnetic potential $A$: $F=dA$. Two potentials
$A,A^{\prime}$ are equivalent iff $dA=dA^{\prime}$; it follows
that $A$ has to be considered as
an operator valued continuous functional on the test space of $\delta$%
-closed (or $\delta$-boundary) $1$-forms, because triviality of
the cohomology implies that $\frac{\Omega_{}^{\prime
1}\left(M\right)}{Z^{d}\Omega_{}^{\prime 1}\left(M\right)}%
=\left(Z^{\delta}\Omega_{c}^{1}\left(M\right)\right)^{\prime}$.
Let $a\in Z^{\delta}\Omega_{c}^{1}\left(M\right)$, then there
exists $f\in \Omega_{c}^{2}\left(M\right)$ such that $a=\delta f$
and $A\left(a\right)=A\left(\delta
f\right)=dA\left(f\right)=F\left(f\right)$ is the electromagnetic
field smeared with the test function $f$.

The antisymmetric bilinear form on $Z^{\delta}\Omega_{c}^{1}\left(M\right)$
or $\Omega_{c}^{2}\left(M\right)$ given by (cf. \cite{LeyRobTesDforQFF})%
\begin{gather*}
\sigma\left(a_{1},a_{2}\right):=\left\langle a_{1},Ea_{2}\right\rangle
=-\int a_{1}\wedge\ast Ea_{2}=-\int\delta f_{1}\wedge\ast E\delta f_{2}
\end{gather*}
is degenerate. We take the quotient of the space of 1-forms with respect to
the degeneracy space, i.e. with respect to the equivalence relation: $a\sim0$
iff $\sigma\left(a,\cdot\right)=0$, or, equivalently, $dEa=0$. We indicate
again with $\sigma$ the induced symplectic form on $Z^{\delta}\Omega_{c}^{1}%
\left(M\right)/\sim$, which becomes in this way a symplectic space. In terms
of the 2-form test function $f$, the symplectic space is $%
\Omega_{c}^{2}\left(M\right)/\sim$, where $f\sim0$ iff $dEf=0$ and $\delta
Ef=0$.

As in the scalar case, we can define in a natural way a complex structure on
$H_{0}:=H\left(\R^{d+1}\right)$ with a square root $J$ of $\mathbf{-1%
}$ given by multiplication in Fourier transform by $i\epsilon\left(p_{0}%
\right)$: $\widehat{Ja}\left(p\right):=i\epsilon\left(p_{0}\right)\widehat{a}%
\left(p\right)$. As $J$ is not a local operator, $H\left(M\right)$ is in
general not invariant under $J$. The propagator $E:\Omega_{c}^{k}\left(%
\R^{d+1}\right)\rightarrow\Omega_{}^{k}\left(\R^{d+1}\right)$%
, when restricted to $\Omega_{c}^{k}\left(M\right)$, gives the
propagator on $M$. The Fock representation corresponds to the
positive semi-definite inner product on $Z^{\delta}\Omega_{c}^{1}\left(%
\R^{d+1}\right)$:
\begin{equation}
\left(a_{1},a_{2}\right)_{\R}:=\left\langle
a_{1},Pa_{2}\right\rangle =-\int a_{1}\wedge\ast Pa_{2}=
-\int\delta f_{1}\wedge\ast P\delta f_{2},
\label{eq:EM-scalar-product}
\end{equation}
where $\ast$ is the Hodge-$\ast$ operator and $P=JE$. The $0$-norm
elements are those equivalent to $0$, therefore the inner product
is positive definite on the quotient space and the completion is
the one-particle complex Hilbert space
\begin{equation}
H_{0}:=H\left(\R^{d+1}\right)=\overline{\frac{Z^{\delta}\Omega_{c}^{1}\left(%
\R^{d+1}\right)}{\sim}}\simeq\overline{\frac{\Omega_{c}^{2}\left(%
\R^{d+1}\right)}{\sim}}\,.  \label{eq:EM-1-phs}
\end{equation}
The symplectic form $\sigma\left(a,b\right)=\left\langle
a,Eb\right\rangle $ can be extended to the closure and the vacuum
state is the Fock state $\omega$ on the CCR-algebra
$\mathfrak{A}\left(H_{0},\sigma\right)$ whose two point function
is the complex scalar product
\begin{equation}
\omega\left(a,b\right)=\left(a,b\right)_{\R}+i\sigma\left(a,b\right)%
\,.  \label{eq:EM-two-point-function}
\end{equation}

\subsection{Cauchy Data}

Let $\cs\subset M\subset\R^{d+1}$ be a Cauchy surface
for $M$ and $\rho_{i}$ the corresponding operators (cf. \ref%
{eq:EM-rho_0-rho_1}). As $M$ is homeomorphic to $\cs \times \R$,
our assumption that $M$ is contractible is equivalent to
contractibility of $\cs$ (and of every Cauchy surface). The
electromagnetic field $F$ and the potential $\left[A\right]$,
which satisfy the wave equation, are determined by their Cauchy data on $%
\cs$ and it is well known that the values of the field and its
normal derivative on $\cs$ are not independent. We now choose to
work with the potential $\left[A\right]\in\frac{\Omega_{}^{\prime
1}\left(M\right)}{Z^{d}\Omega_{}^{\prime 1}\left(M\right)}$ and
the test 1-forms $\left[a\right]$. Given $\left[a\right]\in
Z^{\delta}\Omega_{c}^{1}\left(M\right)/\sim$ then
$\left[Ea\right]\in
Z^{\delta}\Omega_{}^{1}\left(M\right)/\sim$ and $\delta dEa=0$, because on $%
Z^{\delta}\Omega_{}^{1}\left(M\right)$ the equations $\square
a\equiv\left(d\delta+\delta d\right)a=0$ and $\delta da=0$ are
equivalent
and $\square E=0$. The Cauchy data for $Ea$ can be chosen to be $\mathbf{\left[a%
\right]}:=\left[\rho_{0}Ea\right]\in \Omega_{c}^{1}\left(\cs%
\right)/Z^{d}\Omega_{c}^{1}\left(\cs\right),$ where $\mathbf{a}%
\sim0$ iff $d\mathbf{a}=0$, and $\mathbf{e}:=\rho_{1}Ea\in
Z^{\delta}\Omega_{c}^{1}\left(\cs\right)$, as shown in \cite%
{DimEMF}. We quote here the existence statement, \cite{DimEMF} Prop. 2:

\begin{prop}
\label{pro:EM-existence-th}For any $\mathbf{a},\mathbf{e}\in
\Omega_{c}^{1}\left(\cs\right)$ with $\delta\mathbf{e}=0$, there
exists $a\in Z^{\delta}\Omega_{}^{1}\left(M\right)$ such that
$\square a=0$ (i.e. $\delta da=0$), $\rho_{0}a=\mathbf{a}$ and
$\rho_{1}a=\mathbf{e}.$
\end{prop}

The following is a uniqueness statement, \cite{DimEMF} Prop. 3.

\begin{prop}
\label{pro:EM-uniqueness-th}For $i=1,2$, let $a_{i}\in
\Omega_{}^{1}\left(M\right)$ satisfy $\delta da_{i}=0$ and $\rho_{0}a_{i}=%
\mathbf{a}_{i}$, $\rho_{1}a_{i}=\mathbf{e}_{i}$ (so that $\delta\mathbf{e}%
_{i}=0)$, then $a_{1}\sim a_{2}$ iff $\mathbf{e}_{1}=\mathbf{e}_{2}$ and $%
\mathbf{a}_{1}\sim\mathbf{a}_{2}$.
\end{prop}

Notice that the proofs hold in the present context, where our globally
hyperbolic manifold is a subset of Minkowski space-time and we do not
require the Cauchy surface to be compact.

Moreover, like in the scalar case (cf. eq. \ref%
{eq:real-scalar-product-cauchy-data}), considering the effect of change of
time orientation one can conclude that the real scalar product does not
couple the 0-component with the 1-component of the Cauchy data:%
\begin{equation}
\left(a_{1},a_{2}\right)_{\R}=\left\langle
a_{1},Pa_{2}\right\rangle =\left\langle
\mathbf{a}_{1},P_{00}\mathbf{a}_{2}\right\rangle +\left\langle
\mathbf{e}_{1},P_{11}\mathbf{e}_{2}\right\rangle .
\label{eq:EM-scalar-prod-cauchy-data}
\end{equation}

We can then state the following

\begin{thm}
There is an isomorphism of real Hilbert spaces between the space of
(equivalence classes of) covariant potentials and the space of (equivalence
classes of) Cauchy data:%
\begin{equation}
\begin{array}{ccccc}
\overline{\frac{Z^{\delta}\Omega_{c}^{1}\left(M\right)}{\sim}} & \simeq &
\overline{\frac{\Omega_{c}^{1}\left(\cs\right)}{%
Z^{d}\Omega_{c}^{1}\left(\cs\right)}} & \oplus & \overline{%
Z^{\delta}\Omega_{c}^{1}\left(\cs\right)} \\
\left[a\right] & \leftrightarrow & \left[\mathbf{a}\right] & \oplus &
\mathbf{e}%
\end{array}
\label{eq:EM-1PHS-cauchy-data}
\end{equation}
where $\mathbf{a}=\rho_{0}a$ and $\mathbf{e}=\rho_{1}a$ for $a\in \frac{%
Z^{\delta}\Omega_{c}^{1}\left(M\right)}{\sim}$ and the closures are with
respect to the scalar product \ref{eq:EM-scalar-prod-cauchy-data}. The
symplectic form in terms of Cauchy data is
\begin{equation}
\sigma\left(a_{1},a_{2}\right)=\int_{\cs}j^{\ast}\left(Ea_{1}%
\wedge\ast dEa_{2}-Ea_{2}\wedge\ast dEa_{1}\right)=\left\langle \mathbf{a}%
_{1},\mathbf{e}_{2}\right\rangle -\left\langle \mathbf{e}_{1},\mathbf{a}%
_{2}\right\rangle  \label{eq:EM-sympl-form-Cauchy-data}
\end{equation}
\end{thm}

\begin{proof}
The isomorphism is an immediate consequence of Prop. \ref%
{pro:EM-existence-th} and \ref{pro:EM-uniqueness-th} and of the form of the
scalar product \ref{eq:EM-scalar-prod-cauchy-data}.

The expression of the symplectic form
\ref{eq:EM-sympl-form-Cauchy-data} follows by eq.
\ref{eq:EM-E-cauchy-data2} in App. \ref{app:Differential-forms},
noting that the second term in
the sum vanishes when the forms belong to $Z^{\delta}\Omega_{c}^{1}\left(M%
\right)$, because then $\rho_{1}\ast=\ast j^{\ast}\delta$ is zero.
Alternatively, using Stokes' theorem, one can also verify that the integral in \ref%
{eq:EM-sympl-form-Cauchy-data} does not depend on $\cs$ and reduce
the proof of \ref{eq:EM-sympl-form-Cauchy-data} to the known
expression in the case in which $\cs$ is the time-$0$ surface. The
difference of the integrals in \ref{eq:EM-sympl-form-Cauchy-data}
calculated on two (non intersecting) Cauchy surfaces can be
computed as the integral $\int d\left(Ea_{1}\wedge\ast
dEa_{2}-Ea_{2}\wedge\ast dEa_{1}\right)$ on the region enclosed by
the two surfaces; this integral is $0$ because $\delta dEa_{i}=0$.
\end{proof}

\subsection{Local Algebras}

As in the scalar case, the local algebras are associated to the local spaces
defined in the natural way: $H\left(O\right)=\overline{\left[%
Z^{\delta}\Omega_{c}^{1}\left(O\right)\right]}$. Using the Cauchy data, eq. %
\ref{eq:EM-sympl-form-Cauchy-data} becomes for local subspaces
\begin{equation}
H\left(C\left(\mathbf{B}\right)\right)\simeq\mathbf{H}\left(\mathbf{B}%
\right)=\overline{\left[\Omega_{c}^{1}\left(\mathbf{B}\right)\right]\oplus
Z^{\delta}\Omega_{c}^{1}\left(\mathbf{B}\right)}.
\label{eq:EM-isom-cauchy-data}
\end{equation}
The CCR-algebra $\mathfrak{A}\left(H_{0},\sigma\right)$ associated to the
symplectic space $H_{0}$ (eq. \ref%
{eq:EM-1-phs}) is the algebra of observables for the
electromagnetic free field and it has the local structure given by
the local structure of $H_{0}$ in \ref{eq:EM-isom-cauchy-data}:
$\mathfrak{A}\left(O\right):=\mathfrak{A}\left(H\left(O\right)\right)$.

The vacuum state is the state defined by the 2-point function \ref%
{eq:EM-two-point-function}, via the real scalar product \ref%
{eq:EM-scalar-product}, as in the scalar case. In the corresponding
GNS-representation on the Fock space $\Gamma\left(H_{0}\right)$ the weak
closure of the local algebras gives a net of von Neumann algebras $O\mapsto%
\mathfrak{R}\left(O\right)$.

Our main theorem is then:

\begin{thm}
\label{thm:EM-duality-Oc}Let $M$ be an open globally hyperbolic
contractible submanifold of Minkowski space-time $\R^{d+1}$ with
$d\geq2$; for any double cone $O$ with $\overline{O}\subset M$,
the algebra of observables for the free electromagnetic field in
the representation of the vacuum state satisfies relative duality:
\begin{equation*}
\mathfrak{R}\left(O^{c}\right)^{c}=\mathfrak{R}\left(O\right)
\end{equation*}
\end{thm}

The proof will follow from Th. \ref{thm:EM-ops-duality-Oc} below and Lemma %
\ref{lem:duality-ops-duality}, which holds also in this case.

\subsection{One-Particle Space Duality for a Double Cone $O$}

As for $M$, triviality of the cohomology of $\cs$ implies that
\begin{equation} \label{eq:equiv-class-distr-dual-to}
\frac{\Omega_{}^{\prime 1}\left(\cs
\right)}{Z^{d}\Omega_{}^{\prime 1}\left(\cs\right)}=
\left(Z^{\delta}\Omega_{c}^{1}\left(\cs\right)\right)^\prime
\end{equation}
Let us define via the symplectic form a continuous immersion of
the Hilbert space $H$ into a space of (equivalence classes of)
$1$-currents (distribution valued $1$-forms).
\begin{equation*}
\psi:H\supset\frac{\Omega_{c}^{1}\left(\cs\right)}{%
Z^{d}\Omega_{c}^{1}\left(\cs\right)}\oplus
Z^{\delta}\Omega_{c}^{1}\left(\cs\right)\hookrightarrow\frac{%
\Omega_{}^{\prime 1}\left(\cs\right)}{Z^{d}\Omega_{}^{\prime
1}\left(\cs\right)}\oplus Z^{\delta}\Omega_{}^{\prime 1}\left(%
\cs\right).
\end{equation*}
Using eq. \ref{eq:EM-sympl-form-Cauchy-data} one first defines $%
\psi=\psi_{0}\oplus \psi_1$ on $\frac{\Omega_{c}^{1}\left(\cs%
\right)}{Z^{d}\Omega_{c}^{1}\left(\cs\right)}\oplus
Z^{\delta}\Omega_{c}^{1}\left(\cs\right)$ as the identification of
$1$-forms $a_{1}=\left[\mathbf{a}_{1}\right]\oplus \mathbf{e}_{1}$
with
distributions acting on $a_{2}=\left[\mathbf{a}_{2}\right]\oplus \mathbf{e}%
_{2}\in \frac{\Omega_{c}^{1}\left(\cs\right)}{%
Z^{d}\Omega_{c}^{1}\left(\cs\right)}\oplus
Z^{\delta}\Omega_{c}^{1}\left(\cs\right)$ via the canonical
scalar product: it is the direct sum of the maps $\psi_{0}:\frac{%
\Omega_{c}^{1}\left(\cs\right)}{Z^{d}\Omega_{c}^{1}\left(\mathbf{%
\cs}\right)}\hookrightarrow\frac{\Omega_{}^{\prime 1}\left(\cs%
\right)}{Z^{d}\Omega_{}^{\prime 1}\left(\cs\right)}%
=\left(Z^{\delta}\Omega_{c}^{1}\left(\cs\right)\right)^{\prime}$
and $\psi_{1}:Z^{\delta}\Omega_{c}^{1}\left(\cs%
\right)\hookrightarrow Z^{\delta}\Omega_{}^{\prime 1}\left(\cs%
\right)=\left(\frac{\Omega_{c}^{1}\left(\cs\right)}{%
Z^{d}\Omega_{c}^{1}\left(\cs\right)}\right)^{\prime}$, with $%
\left\langle
\psi_{0}\left(\left[a\right]\right),\mathbf{e}\right\rangle
=\left\langle
\mathbf{a},\mathbf{e}\right\rangle $, $\left\langle \psi_{1}\left(\mathbf{e}\right),%
\mathbf{a}\right\rangle =-\left\langle
\mathbf{e},\mathbf{a}\right\rangle $, so that $\left\langle
\psi\left(a_{1}\right),a_{2}\right\rangle =\left\langle
\mathbf{a}_{1},\mathbf{e}_{2}\right\rangle -\left\langle \mathbf{e}_{1},%
\mathbf{a}_{2}\right\rangle $. Then, by the continuity of $\sigma$ with
respect to the norm of $H$, we know that $\psi$ is continuous from $H$ to
the space of distributions and can be extended to the whole Hilbert space.
It is injective because $\sigma$ is non-degenerate and $\frac{%
\Omega_{c}^{1}\left(\cs\right)}{Z^{d}\Omega_{c}^{1}\left(\mathbf{%
\cs}\right)}\oplus Z^{\delta}\Omega_{c}^{1}\left(\cs\right)$ is
dense in $H$.

As in Subsection \ref{sub:Duality-for-Oc}, let $O_{1}\supset\overline{O}$ be
a double cone such that $\overline{O_{1}}\subset M$, choose a Cauchy surface
$\cs$ containing the basis $\mathbf{B}_{1}$ of $O_{1}=C\left(%
\mathbf{B}_{1}\right)\subset\R^{d+1}$ (and the basis $\mathbf{B}$
of $O=C\left(\mathbf{B}\right)$). We prove that
$\mathbf{H}\left(\cs\backslash\overline{\mathbf{B}}\right)^{c}=\mathbf{H}\left(\mathbf{B}\right)$
by proving
eq. \ref{eq:duality-outer-reg} and using outer regularity. Let $a\in \mathbf{%
H}\left(\cs\backslash\overline{\mathbf{B}}\right)^{c}$, then, by
definition of the
relative symplectic complement of $\mathbf{H}\left(\cs\backslash\overline{\mathbf{B}}\right)$, the distribution $\psi\left(a\right)$ has support in $%
\overline{\mathbf{B}}\subset\mathbf{B}_{1}$, where the support of
an equivalence class of distributions is defined in the following
obvious way:

\begin{defn}
The support of an equivalence class of distributions $\left[\mathbf{a}\right]%
\in \frac{\Omega_{}^{\prime 1}\left(\cs\right)}{%
Z^{d}\Omega_{}^{\prime 1}\left(\cs\right)}$ is the
complement of the union of all open sets $\mathbf{A}\subset\cs$ such that $%
Z^{\delta}\Omega_{c}^{1}\left(\mathbf{A}\right)\subset\ker\left[\mathbf{a}%
\right]$ (cf. eq. \ref{eq:equiv-class-distr-dual-to}).
\end{defn}

\begin{prop}
\label{prop:support-a-da} For every $\left[\mathbf{a}\right]\in \frac{%
\Omega_{}^{\prime 1}\left(\cs\right)}{Z^{d}\Omega_{}^{\prime
1}\left(\cs\right)}$, $d\mathbf{a}$ is a well defined $2$%
-current and
\begin{equation*}
\supp \left(d\mathbf{a}\right)=\supp \left[\mathbf{a}\right]=
\bigcap_{\mathbf{a}\in\left[\mathbf{a}\right]}\supp \left(\mathbf{a}%
\right) \, .
\end{equation*}
\end{prop}

\begin{proof}
The support of the distribution $\mathbf{a}$ is the complement of the union
of the open sets $\mathbf{A}$ such that $\mathbf{a}|_\mathbf{A}=0$, where
the last equality means that $\left\langle \mathbf{a},\mathbf{e}%
\right\rangle=0$ for every (not necessarily $\delta$-closed) form
$\mathbf{e} $ with support in $\mathbf{A}$; similarly for
$d\mathbf{a}$. The definition is not altered if the family of open
sets $\mathbf{A}$ is restricted to the open balls.

1. Let $\mathbf{A}$ be an open set, such that $\left\langle \left[\mathbf{a}%
\right],\mathbf{e}\right\rangle=0$ for every $\mathbf{e}\in
Z^{\delta}\Omega_{c}^{1}\left(\mathbf{A}\right)$. For every
$\mathbf{f}\in
\Omega_{c}^{2}\left(\mathbf{A}\right)$ we have $\left\langle \left[\mathbf{a}%
\right],\mathbf{\delta f}\right\rangle=\left\langle d\mathbf{a},\mathbf{f}%
\right\rangle=0$, thus $d\mathbf{a}|_\mathbf{A}=0$. It follows
that $\supp \left(d\mathbf{a}\right) \subset \supp
\left[\mathbf{a}\right]$.

2. Let $\mathbf{A}$ be an open set, if there is an $\mathbf{a}\in\left[%
\mathbf{a}\right]$ such that $\mathbf{a}|_\mathbf{A}=0$, then $\left[\mathbf{%
a}\right]|_\mathbf{A}=0$. This implies that $\supp \left[\mathbf{a}%
\right]\subset \bigcap_{\mathbf{a}\in\left[\mathbf{a}\right]}\supp
\left(\mathbf{a}\right)$.

3. Let $\mathbf{A}$ be an open ball in $\R^d$, such that $d\mathbf{a}%
|_\mathbf{A}=0$. As $\mathbf{A}$ is contractible, there is a
distribution $\phi:\mathbf{A}\rightarrow\R$ such that $\mathbf{a}|_\mathbf{A}%
=d\phi$. A distribution cannot in general be extended to $\R^d$.
Nevertheless, for any open ball $\mathbf{A_1}$ such that
$\overline{\mathbf{A_1}}\subset\mathbf{A}$, there exists a
continuous function $\varphi \in
C^0\left(\overline{\mathbf{A_1}}\right)$ such that
$\phi|_{\mathbf{A_1}}$ is the $n$-th derivative of $\varphi$,
$\phi|_{\mathbf{A_1}} =
\partial^n \varphi|_{\mathbf{A_1}}$. $\varphi$ can be extended to a (not
necessarily continuous) function on $\R^d$. $\phi^\prime =
\partial^n \varphi$ is a distribution on $\R^d$ that coincides
with $\phi$
on $\mathbf{A_1}$, thus $\mathbf{a}_1:=\mathbf{%
a}-d\phi^\prime$ is such that
$\left[\mathbf{a}_1\right]=\left[\mathbf{a}\right]$ and
$\mathbf{a}_1|_\mathbf{A_1}=0$. Therefore $\mathbf{A_1} \subset
\bigcup_{\mathbf{a}\in\left[\mathbf{a}\right]}\left(\supp
\left(\mathbf{a}\right)\right)^c$ whenever
$\overline{\mathbf{A_1}}
\subset \mathbf{A}$. This implies that $\bigcap_{\mathbf{a}\in%
\left[\mathbf{a}\right]}\supp \left(\mathbf{a}\right) \subset
\supp \left(d\mathbf{a}\right)$.

\end{proof}

\begin{rem}
\label{rem:cohomology}It is a cohomological problem to know if
$\supp \left[\mathbf{a}\right]=\mathbf{K}\subset\mathbf{B}$, with
$\mathbf{B}$ open, implies that there is a representative
$\mathbf{a}\in
\Omega_{}^{\prime 1}\left(\mathbf{B}\right)$. We have that $\supp \left[%
\mathbf{a}\right]\subset\mathbf{B}$ implies $\supp \left(d\mathbf{a}\right)\subset%
\mathbf{B}$ and, if $d\mathbf{a}$ is trivial in the second cohomology group $H^{2}\left(\mathbf{B}%
\right)$ of $\mathbf{B}$, then $d\mathbf{a}\in
B^{d}\Omega_{}^{\prime 2}\left(\mathbf{B}\right)$; this means that
there exists $\tilde{\mathbf{a}}\in \Omega_{}^{\prime
1}\left(\mathbf{B}\right)$
such that $d\mathbf{a}=d\tilde{\mathbf{a}}$, which implies that $\left[%
\mathbf{\tilde{\mathbf{a}}}\right]=\left[\mathbf{a}\right]$.
\end{rem}

When we restrict our attention to the double cone $O_{1}$, the Cauchy surface $%
\cs$ can be identified with $\R^{d}$ and we can use the known
results on Minkowski space-time and time-$0$ Cauchy surface,
including outer regularity \cite{LeyRobTesDforQFF}. To find
regular forms approximating $a\in
\mathbf{H}\left(\cs\backslash\overline{\mathbf{B}}\right)^{c}$ we
use convolutions with regular functions $\rho_{n}$ approximating
$\delta$.

\begin{prop}
\label{pro:EM-Crho}Let $\rho,\rho_{n}\in C_{c}^{\infty}\left(\R%
^{d}\right)$ be as in Subsection \ref{sub:Duality-for-Oc}, then there exists
a sequence of convolution operators with $\rho_{n}$
\begin{equation*}
C_{n}:\frac{\Omega_{c}^{\prime 1}\left(\cs\right)}{%
Z^{d}\Omega_{c}^{\prime 1}\left(\cs\right)}\oplus
Z^{\delta}\Omega_{c}^{\prime 1}\left(\cs\right)\rightarrow\frac{%
\Omega_{c}^{1}\left(\cs\right)}{Z^{d}\Omega_{c}^{1}\left(\mathbf{%
\cs}\right)}\oplus Z^{\delta}\Omega_{c}^{1}\left(\cs\right)
\end{equation*}
such that, for any choice of relatively compact open sets $\mathbf{B}\subset%
\overline{\mathbf{B}}\subset\mathbf{B}_{1}\subset\cs$, $\supp
\alpha\subset\mathbf{B}$ implies $\supp C_{n}\alpha\subset\mathbf{%
B}_{1}$ for large $n$ and $C_{n}\rightarrow\mathbf{1}$ strongly when
restricted to $\mathbf{H}\left(\mathbf{B}\right)$.
\end{prop}

\begin{proof}
Let $\rho,\rho_{n}\in C_{c}^{\infty}\left(\R^{d}\right)$ be as in
Subsection \ref{sub:Duality-for-Oc}, consider the convolution operator $%
C_{\rho_{n}}:\Omega_{c}^{\prime k}\left(\cs\right)\rightarrow%
\Omega_{c}^{k}\left(\cs\right)$. It commutes with $d$ and $%
\delta,$ therefore the image of $Z^{d}\Omega_{}^{\prime 1}\left(\mathbf{%
\cs}\right)$ is in $Z^{d}\Omega_{c}^{1}\left(\cs\right)$ and the
image of $Z^{\delta}\Omega_{}^{\prime 1}\left(\cs\right)$ is
in $Z^{\delta}\Omega_{c}^{1}\left(\cs\right)$, so that $%
C_{\rho_{n}}$ induces an operator $\widetilde{C_{\rho_{n}}}$ on the quotient
space and we define $C_{n}:=\widetilde{C_{\rho_{n}}}\oplus C_{\rho_{n}}$.

In the double cone $O_{1}=C\left(\mathbf{B}_{1}\right)$, the Cauchy surface $%
\cs$ can be identified with $\R^{d}$ and using Fourier
transform (see \cite{LeyRobTesDforQFF}) we prove as in the scalar case that $%
C_{n}a\rightarrow a$ in $\mathbf{H}\left(\mathbf{B}\right)$.
\end{proof}

\begin{thm}
\label{thm:EM-ops-duality-Oc}For any double cone $O$ such that $\overline{O}%
\subset M$ with basis $\mathbf{B}$, one-particle space duality \ref%
{eq:one-particle-space-duality-Oc} is satisfied:
\begin{equation}
H\left({O}^{c}\right)^{c}=H\left(O\right)\,,\quad\quad\quad\mathbf{H}
\left(\cs\backslash\overline{\mathbf{B}}\right)^{c}=\mathbf{H}\left(\mathbf{B}\right)\,.
\label{eq:EM-1-p-s-duality}
\end{equation}
\end{thm}

\begin{proof}
For any $a\in
\mathbf{H}\left(\cs\backslash\overline{\mathbf{B}}\right)^{c}$ we
want to show that $a\in \mathbf{H}\left(\mathbf{B}\right)$. The forms $%
a_{n}:=C_{n}\psi\left(a\right)=\left[\mathbf{a}_{n}\right]\oplus \mathbf{e}%
_{n}$ are in $\frac{\Omega_{c}^{1}\left(\cs\right)}{%
Z^{d}\Omega_{c}^{1}\left(\cs\right)}\oplus
Z^{\delta}\Omega_{c}^{1}\left(\cs\right)$ and for any open set $\mathbf{B}_{1}\supset\overline{%
\mathbf{B}}$ have support in $\mathbf{B}_{1}$ for large $n$,
because $\supp \left(a_{n}\right) \subset \supp
\left(\psi\left(a\right)\right)+\supp \left(\rho_{n}\right)$.
We have that $\mathbf{e}_{n}\in \Omega_{}^{\prime 1}\left(%
\mathbf{B}_{1}\right)$ and, by the Remark \ref{rem:cohomology}, as $\mathbf{B%
}$ is contractible and has trivial 2-cohomology, also
$\mathbf{a}_{n}$ can
be chosen in $\Omega_{}^{\prime 1}\left(\mathbf{B}_{1}\right)$, therefore $%
a_{n}\in \mathbf{H}\left(\mathbf{B_{1}}\right)$. Finally, $a=\lim a_{n}$ is
in $\bigcap_{\mathbf{B_{1}\supset}\overline{\mathbf{B}}}\mathbf{H}\left(%
\mathbf{B_{1}}\right)$ and, by outer regularity and Prop. \ref{pro:EM-Crho},
$a\in \mathbf{H}\left(\mathbf{B}\right)$.
\end{proof}

\subsection{A Special Case of Relative Duality in $V_{+}$}

We are now more specific and consider the forward light cone
$V_{+}$ and as a Cauchy surface $\cs$ the positive time branch of
the hyperboloid $x^{2}=c^2$, $c\in \R_{+}$. Let
$\mathcal{C}\subset \R^d$ be an open cone around a specific
(space-like) direction $\mathbf{v}\in\R^d$:
\begin{equation} \label{eq:cone}
\mathcal{C}=\left\{\mathbf{x}\in\R^d :
\left(\mathbf{x},\mathbf{v}\right)>\left( 1-\epsilon \right)
\left\Vert x \right\Vert \right\}
\end{equation}
with $\mathbf{v}^2=1$ and small $\epsilon>0$. Let
$\mathbf{A}\subset\cs$ be the open set of points obtained by
acting on the point $\left(c,\mathbf{0}\right)$ with the
semi-group of boosts with speed belonging to $\mathcal{C}$.
$\mathbf{A}$ is an unbounded open set lying on $\cs$ with vertex
in $\left(c,\mathbf{0}\right)$ and escaping to light-like infinity
along the direction $\mathbf{v}$. This is physically interesting
(see Introduction) as a state with an electric charge in $V_{+}$
can be obtained as a limit of neutral states with another
compensating charge that is moved to time-like infinity along
$\mathbf{A}$. Such a state would be localized in $\mathbf{A}$ or,
equivalently, in its causal completion
$O:=C\left(\mathbf{A}\right)$.

We want to prove that relative duality
holds for $\mathbf{H}\left(\mathbf{A}\right)$: $\mathbf{H}\left({\cs\backslash%
\overline{\mathbf{A}}}\right)^{c}=\mathbf{H}\left(\mathbf{A}\right)$.

As a first step we use conformal covariance to reduce the problem
to a simpler situation in which the ambient space is a double
cone. We consider a
conformal map $\varphi$ as in Subsection \ref{sub:massless-case} such that $%
O_{1}:=\varphi\left(V_{+}\right)$ is a double cone with basis $\mathbf{B}%
_{1}:=\varphi\left(\cs\right)$ lying on the time-$0$ surface $%
\R^{d}\subset\R^{d+1}$. $\mathbf{B}:=\varphi\left(\mathbf{A}%
\right)\subset\mathbf{B}_{1}$ is a bounded open strongly contractible subset of $%
\R^{d}$ (the intersection of a cone and $\mathbf{B}_{1}$),
the causal completion of $\mathbf{B}$ is $C\left(\mathbf{B}%
\right)=C\left(\varphi\left(\mathbf{A}\right)\right)=\varphi\left(C\left(%
\mathbf{A}\right)\right)=\varphi\left(O\right)$ and the relative causal
complement in $O_{1}$ is $\mathbf{B}^{c}=C\left(\mathbf{B}%
\right)^{\prime}\cap
O_{1}=\varphi\left(O\right)^{\prime}\cap\varphi\left(V_{+}\right)=\varphi%
\left(O^{\prime}\cap V_{+}\right)=\varphi\left(O^{c}\right)$. As
the free electromagnetic field is massless, it is conformally
covariant, i.e. there is
a unitary implementation $U\left(\varphi\right)$ of the conformal map $%
\varphi$ (cf. \cite{MR89j:81088}): $U\left(\varphi\right)^{\ast}\mathfrak{R}%
\left(O\right)U\left(\varphi\right)=\mathfrak{R}\left(\varphi\left(O\right)%
\right)$ for any $O\subset V_{+}$, therefore $\mathfrak{R}%
\left(O^{\prime}\cap V_{+}\right)^{\prime}\cap\mathfrak{R}\left(V_{+}\right)=%
\mathfrak{R}\left(O\right)$ $\iff$ $\mathfrak{R}\left(\varphi\left(O\right)^{%
\prime}\cap\varphi\left(V_{+}\right)\right)^{\prime}\cap\mathfrak{R}%
\left(\varphi\left(V_{+}\right)\right)=\mathfrak{R}\left(\varphi\left(O%
\right)\right)$ $\iff$
$\mathfrak{R}\left(\mathbf{B}_{1}\backslash\overline{\mathbf{B}}
\right)^{c}=\mathfrak{R}\left(\mathbf{B}\right)$ and again this
property can be reduced to the one-particle space duality, namely
\begin{equation} \label{eq:1-part-space-duality-B}
\mathbf{H}\left(\mathbf{B}_{1}\backslash\overline{\mathbf{B}}%
\right)^{c}=\mathbf{H}\left(%
\mathbf{B}\right) \, ,
\end{equation}
for a strongly contractible subset $\mathbf{B}\subset%
\mathbf{B}_{1}\subset\R^{d}$.

The one-particle Hilbert space $H_{0}=H\left(\R^{d+1}\right)$,
written in terms of Cauchy data on the standard time-$0$ Cauchy
surface, can
be described as in \ref{eq:EM-1PHS-cauchy-data} and the scalar product in %
\ref{eq:EM-scalar-prod-cauchy-data} (or the corresponding norm)
can be
explicitly written in this special case using Fourier transform: $\left\Vert %
\left[\mathbf{a}\right]\oplus \mathbf{e}\right\Vert ^{2}=\left\Vert P_{T}%
\mathbf{a}\right\Vert _{+}^{2}+\left\Vert \mathbf{e}\right\Vert _{-}^{2}$,
where $P_{T}$ is the projection on the transverse (or divergence free) part,
$\widehat{P_{T}a}_{i}\left(\mathbf{p}\right)= \left(\delta_{i,j}-{p_{i}p_{j}}%
/{\mathbf{p}^{2}}\right)\widehat{a}_{j}\left(\mathbf{p}\right)$, and the
norms used are the usual free field norms $\left\Vert \mathbf{f}\right\Vert
_{\pm}^{2}= \int\left\Vert \widehat{\mathbf{f}}\left(\mathbf{p}%
\right)\right\Vert ^{2}\left\Vert \mathbf{p}\right\Vert ^{\pm1}d^{d}\mathbf{p%
}$. There is a one to one correspondence between the vector potential $\left[%
\mathbf{a}\right]$ and the magnetic field $\mathbf{b}=\ast d \mathbf{a}$. In
terms of $\mathbf{b}$ and $\mathbf{e}$ the one particle space norm becomes
(cf. \cite{BenNicLAforSFF&EMFF})
\begin{equation} \label{eq:VF-norm-cauchy-data}
\left\Vert \mathbf{b}\oplus \mathbf{e}\right\Vert ^{2}=\left\Vert \mathbf{b}%
\right\Vert _{-}^{2}+\left\Vert \mathbf{e}\right\Vert _{-}^{2}\,.
\end{equation}

We need a simple lemma which states that there are no elements in
the Hilbert space with support on the boundary of a sufficiently
regular region.

\begin{lem}
\label{lem:regular-boundary}Let $\mathbf{B}$ be an open bounded
strongly contractible set such that
$\mathbf{H}\left(\mathbf{B}\right)$ is outer regular, then there
are no non zero elements $a\in H$ such that $\supp
\psi\left(a\right)\subset\partial\mathbf{B}$.
\end{lem}

\begin{proof}
Suppose $a\in H$ and $\supp \psi\left(a\right)\subset\partial\mathbf{B}$%
, then by a similar reasoning as in Th.
\ref{thm:EM-ops-duality-Oc} above,
we can conclude that $a\in \bigcap_{\mathbf{B_{1}\supset}\overline{\mathbf{B}%
}}\mathbf{H}\left(\mathbf{B_{1}}\right)$ (notice that
$\partial\mathbf{B}$ in general is not contractible but
$\mathbf{B}$ is and the sets $\mathbf{B}_1$ can be
chosen to be). On the other side, $\supp \psi\left(a\right)\subset%
\partial\mathbf{B}$ implies that $\sigma\left(a,a_{1}\right)=0$ for any $%
a_{1}\in \left[\Omega_{c}^{1}\left(\mathbf{B}\right)\right]\oplus
Z^{\delta}\Omega_{c}^{1}\left(\mathbf{B}\right)$ and, by continuity of $%
\sigma$, for any $a_{1}\in \mathbf{H}\left(\mathbf{B}\right)$,
i.e. $a\in
\mathbf{H}\left(\mathbf{B}\right)^{\prime}$. If $\mathbf{H}\left(\mathbf{B}%
\right)$ is outer regular, then $a\in \mathbf{H}\left(\mathbf{B}\right)\cap%
\mathbf{H}\left(\mathbf{B}\right)^{\prime}$ and, as $\sigma$ is
non-degenerate on $\mathbf{H}\left(\mathbf{B}\right)$, $a=0$.
\end{proof}

In view of Lemma \ref{lem:support-boundary}, which generalizes
Lemma \ref{lem:regular-boundary}, we prove the following:

\begin{lem}
\label{lem:MultOperator}For every function $\chi \in C_{c}^{\infty }\left(
\R^{d}\right) $, the multiplication operator $M_{\chi }$ on $%
\Omega _{c}^{1}\left( \R^{d}\right) \oplus \Omega _{c}^{1}\left(
\R^{d}\right) $ with the norm \ref{eq:VF-norm-cauchy-data} is
continuous.
\end{lem}

\begin{proof}
In Fourier transform, the multiplication operator $M_{\chi }$
becomes the convolution operator $T_{\widetilde{\chi}}$ with
kernel $\widetilde{\chi }$ and, according to
\ref{eq:VF-norm-cauchy-data}, the norm to be used is $\left\Vert
\omega^{- \frac{1}{2}}\widetilde{\mathbf{b}}\right\Vert_2$, where
$\omega$ is the multiplication operator by the function
$\omega\left(\mathbf{p}\right)=\left\Vert\mathbf{p}\right\Vert$.
We have to prove that there exists a constant $C$ such that
$\left\Vert \omega^{- \frac{1}{2}} T_{\widetilde{\chi}}
\widetilde{\mathbf{b}}\right\Vert_2 \leq C \left\Vert \omega^{-
\frac{1}{2}} \widetilde{\mathbf{b}}\right\Vert_2$ and this follows
if we prove that the operator
\begin{equation} \label{eq:operator-mult-chi-mom}
\omega^{\pm \frac{1}{2}}T_{\widetilde{\chi }} \omega^{\mp
\frac{1}{2}}
\end{equation}%
is bounded on $%
L^{2}\left( \R^{d}\right)$. We consider separately the infrared
and ultraviolet behavior of the operator. Let $P_{\left[ 0,1\right] }$ and $%
P_{\left( 1,\infty \right) }=\mathbf{1}-P_{\left[ 0,1\right] }$ be
the spectral operators associated to $\omega \left(
\mathbf{p}\right) \leq 1$ and $\omega
\left( \mathbf{p}\right) >1$, then we decompose the operator \ref%
{eq:operator-mult-chi-mom} as the sum of $P_{\left[ 0,1\right] }\omega ^{\pm
\frac{1}{2}}T_{\widetilde{\chi }}\omega ^{\mp \frac{1}{2}}P_{\left[ 0,1%
\right] }$, $P_{\left[ 0,1\right] }\omega ^{\pm \frac{1}{2}}T_{\widetilde{%
\chi }}\omega ^{\mp \frac{1}{2}}P_{\left( 1,\infty \right) }$, $P_{\left(
1,\infty \right) }\omega ^{\pm \frac{1}{2}}T_{\widetilde{\chi }}\omega ^{\mp
\frac{1}{2}}P_{\left[ 0,1\right] }$, $P_{\left( 1,\infty \right) }\omega
^{\pm \frac{1}{2}}T_{\widetilde{\chi }}\omega ^{\mp \frac{1}{2}}P_{\left(
1,\infty \right) }$ and prove that they are bounded.

First of all, $T_{\widetilde{\chi }}$ is bounded by Schur's test: using
different estimates for small and for large $\mathbf{p}-\mathbf{q}$, there exist constants $%
C_{n}$ such that $\left\vert \widetilde{\chi }\left(
\mathbf{p}-\mathbf{q}\right) \right\vert
\leq C_{n}\left\Vert \mathbf{p}-\mathbf{q}\right\Vert ^{-n}$, therefore%
\begin{multline*}
\sup_{\mathbf{p}}\int \left\vert \widetilde{\chi }\left(
\mathbf{p}-\mathbf{q}\right) \right\vert d\mathbf{q}\leq
\sup_{\mathbf{p}}\left\{ \int_{\left\Vert
\mathbf{p}-\mathbf{q}\right\Vert \leq
1}C_{0}d\mathbf{q}+\int_{\left\Vert
\mathbf{p}-\mathbf{q}\right\Vert
>1}C_{n}\left\Vert \mathbf{p}-\mathbf{q}\right\Vert
^{-n}d\mathbf{q}\right\}  \\
=\sup_{\mathbf{p}}\left\{ \int_{\left\Vert \mathbf{q}\right\Vert
\leq 1}C_{0}d\mathbf{q}+\int_{\left\Vert \mathbf{q}\right\Vert
>1}C_{n}\left\Vert \mathbf{q}\right\Vert ^{-n}d\mathbf{q}\right\} <+\infty
\end{multline*}%
and the same holds when $\mathbf{p}$ and $\mathbf{q}$ are
interchanged. Concerning the first
three operators, we prove that $\left( \omega ^{\pm \frac{1}{2}}T_{%
\widetilde{\chi }}\omega ^{\mp \frac{1}{2}}-T_{\widetilde{\chi }}\right) P_{%
\left[ 0,1\right] }$ is Hilbert-Schmidt, it then follows that $\omega ^{\pm \frac{1}{2}}T_{%
\widetilde{\chi }}\omega ^{\mp \frac{1}{2}}P_{\left[ 0,1\right] }$ is
bounded and so are its adjoint $P_{\left[ 0,1\right] }\omega ^{\mp \frac{1}{2}}T_{%
\widetilde{\chi }}\omega ^{\pm \frac{1}{2}}$ and the three
operators. We use the estimate
\begin{equation}
\left\vert \frac{\left\Vert \mathbf{p}\right\Vert
^{\frac{1}{2}}-\left\Vert
\mathbf{q}\right\Vert ^{\frac{1}{2}}}{\left\Vert \mathbf{q}\right\Vert ^{\frac{1}{2}}}%
\right\vert =\left\vert \frac{\left\Vert \mathbf{p}\right\Vert
-\left\Vert \mathbf{q}\right\Vert }{\left\Vert
\mathbf{q}\right\Vert ^{\frac{1}{2}}\left( \left\Vert
\mathbf{p}\right\Vert ^{\frac{1}{2}}+\left\Vert \mathbf{q}\right\Vert ^{\frac{1}{2}}\right) }%
\right\vert \leq \frac{\left\Vert \mathbf{p}-\mathbf{q}\right\Vert }{\left\Vert \mathbf{q}\right\Vert ^{%
\frac{1}{2}}\left\Vert \mathbf{p}\right\Vert ^{\frac{1}{2}}}
\label{eq:estimate-w1/2-w1/2}
\end{equation}%
which is a symmetric expression and therefore holds also when
$\mathbf{p}$ and $\mathbf{q}$
are interchanged. The kernel of the operator $\left( \omega ^{\pm \frac{1}{2}%
}T_{\widetilde{\chi }}\omega ^{\mp \frac{1}{2}}-T_{\widetilde{\chi
}}\right) P_{\left[ 0,1\right] }$ is square integrable,
essentially because the integral of $\widetilde{\chi }\left(
\mathbf{p}-\mathbf{q}\right) $ in $\mathbf{q}$ over a compact
subset gives a rapid decreasing function of $\mathbf{p}$:
\begin{multline*}
\int \int_{\left\Vert \mathbf{q}\right\Vert \leq 1}\left\vert
\frac{\left\Vert
\mathbf{p}\right\Vert ^{\frac{1}{2}}-\left\Vert \mathbf{q}\right\Vert ^{\frac{1}{2}}}{%
\left\Vert \mathbf{q}\right\Vert ^{\frac{1}{2}}}\widetilde{\chi
}\left( \mathbf{p},\mathbf{q}\right) \right\vert ^{2}
d\mathbf{q}\,d\mathbf{p}
\leq
\int \left( \int_{\left\Vert
\mathbf{q}\right\Vert \leq 1}\left\Vert
\mathbf{p}-\mathbf{q}\right\Vert ^{2}\left\vert \widetilde{\chi
}\left(
\mathbf{p}-\mathbf{q}\right) \right\vert ^{2}\frac{d\mathbf{q}}{\left\Vert \mathbf{q}\right\Vert }\right) \frac{%
d\mathbf{p}}{\left\Vert \mathbf{p}\right\Vert }
\leq \\%
\leq \int_{\left\Vert \mathbf{p}\right\Vert \leq
2}\int_{\left\Vert \mathbf{q}\right\Vert \leq
1}C_{1}^{2}\frac{d\mathbf{q}}{\left\Vert \mathbf{q}\right\Vert
}\frac{d\mathbf{p}}{\left\Vert \mathbf{p}\right\Vert } +
\int_{\left\Vert \mathbf{p}\right\Vert
>2}\left( \int_{\left\Vert
\mathbf{q}\right\Vert \leq 1}C_{n+1}^{2}\left\Vert \mathbf{p}-\mathbf{q}\right\Vert ^{-2n}\frac{d\mathbf{q}}{%
\left\Vert \mathbf{q}\right\Vert }\right)
\frac{d\mathbf{p}}{\left\Vert \mathbf{p}\right\Vert }<+\infty
\end{multline*}%
The operator is thus Hilbert-Schmidt.

Finally, the operator $P_{\left( 1,\infty \right) } \left( \omega
^{\pm \frac{1}{2}}T_{\widetilde{\chi }}\omega ^{\mp \frac{1}{2}} -
T_{\widetilde{\chi }} \right) P_{\left( 1,\infty
\right) }$ is bounded, again by Schur's test: using the estimate \ref%
{eq:estimate-w1/2-w1/2}, we have that
\begin{multline*}
\sup_{\left\Vert \mathbf{p} \right\Vert \geq 1} \int_{\left\Vert
\mathbf{q}\right\Vert \geq 1}\left\vert \frac{\left\Vert
\mathbf{p}\right\Vert ^{\frac{1}{2}}-\left\Vert \mathbf{q}\right\Vert ^{\frac{1}{2}}}{%
\left\Vert \mathbf{q}\right\Vert ^{\frac{1}{2}}}\widetilde{\chi
}\left( \mathbf{p},\mathbf{q}\right) \right\vert  d\mathbf{q} \leq
\sup_{\left\Vert \mathbf{p} \right\Vert \geq 1} \int_{\left\Vert
\mathbf{q}\right\Vert \geq 1}\left\Vert
\mathbf{p}-\mathbf{q}\right\Vert \left\vert \widetilde{\chi
}\left(
\mathbf{p}-\mathbf{q}\right) \right\vert d\mathbf{q}\leq  \\
\leq \sup_{\mathbf{p}}\left\{ \int_{\left\Vert
\mathbf{p}-\mathbf{q}\right\Vert \leq
1}C_{1}d\mathbf{q}+\int_{\left\Vert
\mathbf{p}-\mathbf{q}\right\Vert >1}C_{n+1}\left\Vert
\mathbf{p}-\mathbf{q}\right\Vert ^{-n}d\mathbf{q}\right\} <+\infty
\end{multline*}%
and by symmetry the same holds when $\mathbf{p}$ and $\mathbf{q}$
are interchanged.
\end{proof}

\begin{lem}
\label{lem:support-boundary}Let $\mathbf{B}_{1}$ be an open
bounded contractible set
such that $\mathbf{H}\left(\mathbf{B_{1}}\right)$ is outer regular, let $%
a\in \mathbf{H}\left(\mathbf{B_{1}}\right)$ be such that $\supp
\psi\left(a\right)\subset\overline{\mathbf{B}}\cup\partial\mathbf{B}_{1}$,
with $\mathbf{B}\subset\mathbf{B}_{1}$ open set, then $\supp
\psi\left(a\right)\subset\overline{\mathbf{B}}$.
\end{lem}

\begin{proof}
In order to apply Lemma \ref{lem:regular-boundary} we need to decompose $a$
as a sum of two elements, one with support in $\overline{\mathbf{B}}$ and
the other with support in $\partial\mathbf{B}_{1}$. Such a decomposition can
be obtained by using a multiplication operator $M_{\chi}$ with a $\chi\in
C_{c}^{\infty}\left(\R^{d}\right)$ such that $\chi\equiv1$ on $%
\overline{\mathbf{B}}$ and $\supp
\chi\subset\mathbf{B}_{\epsilon}$,
where $\mathbf{B}_{\epsilon}:=\mathbf{B}+\left\{ x\in \R%
^{d}:\left\Vert x\right\Vert <\epsilon\right\} $. The space $H_{0}$ is a
space of divergence free vector fields and is therefore not invariant under
multiplication by a function. However, $H_{0}{}$ is the 'transverse' part of
the Hilbert space $K$ defined as the completion of the space of couples of
vector fields $\mathbf{e},\mathbf{b}\in \Omega_{c}^{1}\left(\R%
^{d}\right)$ with respect to the norm \ref{eq:VF-norm-cauchy-data}. By Prop. %
\ref{prop:support-a-da}, $\mathbf{e}$ and $\mathbf{b}$ are distributions
with support in $\overline{\mathbf{B}}\cup\partial\mathbf{B}_{1}$ belonging
to $K$.

By Lemma \ref{lem:MultOperator}, the continuous operator
$M_{\chi}$ can be extended from
$\Omega_{c}^{1}\left(\R^{d}\right)\oplus
\Omega_{c}^{1}\left(\R^{d}\right)$  to the closure $K$. $M_{\chi}\mathbf{b}$ and $\left(\mathbf{1}%
-M_{\chi}\right)\mathbf{b}=M_{1-\chi}\mathbf{b}$ are in $K$,
$\supp \left(M_{\chi}\mathbf{b}\right)\subset\supp \chi\cap\supp \mathbf{%
b}\subset\mathbf{B}_{\epsilon}$, $\supp \left(M_{1-\chi}\mathbf{b}%
\right)\subset\supp \left(1-\chi\right)\cap\supp \mathbf{b}%
\subset\partial\mathbf{B}_{1}$. $M_{1-\chi}\mathbf{b}$ is a
distribution with support in $\partial\mathbf{B}_{1}$, but belongs
also to a space of free massless vector fields, which are a direct
sum of free scalar fields. Therefore, it is outer regular and
Lemma
\ref{lem:regular-boundary} can be applied to conclude that $M_{1-\chi}\mathbf{b}=0$, thus $\mathbf{%
b}=M_{\chi}\mathbf{b}$ and $\supp \left(\mathbf{b}\right)\subset\mathbf{%
B}_{\epsilon}$. Similarly for $\mathbf{e}$. It follows that $a$,
as a distribution $\psi\left(a\right)$, has support in
$\mathbf{B}_{\epsilon}$ for
$\epsilon$ arbitrary, therefore $\supp \psi\left(a\right)%
\subset\overline{\mathbf{B}}$.
\end{proof}

We can now prove the one-particle space duality for the considered regions.

\begin{thm} \label{thm:EM-ops-duality-A}
Let $\cs=\left\{ x\in\R%
^{d+1} : x_0\geq 0, x^2=c\in \R_+ \right\}$ be an hyperboloid in
the forward light cone $V_{+}\subset\mst$ and $\mathcal{C}\subset
\R^d$ an open cone around a specific (space-like) direction (eq.
\ref{eq:cone}); let $\mathbf{A}\subset\cs$ be the open set of
points obtained by acting on the point $\left(c,\mathbf{0}\right)$
with the semi-group of boosts with speed belonging to
$\mathcal{C}$ and $C\left(\mathbf{A}\right)\subset V_{+}$ its
causal completion. Then,
the isomorphic spaces $H\left(C\left(\mathbf{A}\right)\right) \simeq \mathbf{H%
}\left(\mathbf{A}\right)$ satisfy one-particle space duality in
$H\left(V_+\right) \simeq \mathbf{H}\left(\cs\right)$:
\begin{equation*}
H\left({C\left(\mathbf{A}\right)}^{c}\right)^{c}=H\left(C\left(\mathbf{A}%
\right)\right)\,,\quad\quad\quad
\mathbf{H}\left({\cs\backslash\overline{\mathbf{A}}}\right)^{c}=\mathbf{H}\left(\mathbf{A}\right)\,.
\end{equation*}
\end{thm}

\begin{proof}
As explained above, by conformal covariance the previous equation
is equivalent to eq. \ref{eq:1-part-space-duality-B} for a
strongly contractible subset $\mathbf{B}$ of an open ball
$\mathbf{B}_1 \subset \R^d$. Using dilations and the fact that
$\mathbf{B}$ is strongly contractible (cf. App.
\ref{app:Outer-Regularity} and \cite{LeyRobTesDforQFF}), it can be
proved that the subspace $\mathbf{H}\left(\mathbf{B}\right)$ is
outer regular.

Let $a\in \mathbf{H}\left(\mathbf{B}_1\backslash\overline{\mathbf{B}}\right)^{c}$, then $%
a\in \mathbf{H}\left(\mathbf{B_{1}}\right)$ implies
$\supp \psi\left(a\right)\subset\overline{\mathbf{B}_{1}}$ and $%
a\in \mathbf{H}\left(\mathbf{B_{1}\backslash\overline{B}}%
\right)^{\prime}$ implies $\supp \psi\left(a\right)\subset\left(\mathbf{B}%
_{1}\backslash\overline{\mathbf{B}}\right)^{c}$, hence $\supp
\psi\left(a\right)\subset\overline{\mathbf{B}}\cup\partial\mathbf{B}_{1}$
and, by Lemma \ref{lem:support-boundary}, $\supp \psi\left(a\right)%
\subset\overline{\mathbf{B}}$.

We can now use the same argument as in Th.
\ref{thm:EM-ops-duality-Oc} to
define $C_{n}\psi\left(a\right) \in \frac{\Omega_{c}^{1}\left(\mathbf{B}+\mathbf{%
B}_{\epsilon}\right)}{Z^{d}\Omega_{c}^{1}\left(\cs\right)}\oplus
Z^{\delta}\Omega_{c}^{1}\left(\mathbf{B}+\mathbf{B}_{\epsilon}\right)$
and conclude, using outer regularity of
$\mathbf{H}\left(\mathbf{B}\right)$, that $a=\lim C_{n}a\in
\mathbf{H}\left(\mathbf{B}\right)$.
\end{proof}

It then follows the main result of this Subsection:

\begin{thm} \label{thm:EM-rel-duality-A}
Let $O:=C\left(\mathbf{A}\right)$ in the ambient space $V_+$ as in
Th. \ref{thm:EM-ops-duality-A}, then the algebra of observables
for the free electromagnetic field in the representation of the
vacuum state satisfies relative duality:
\begin{equation*}
\mathfrak{R}\left(O^{c}\right)^{c}=\mathfrak{R}\left(O\right) \,.
\end{equation*}
\end{thm}

\appendix

\section{\label{app:Outer-Regularity}Outer Regularity}

Outer regularity (Def. \ref{def:outer-regularity}) for $H\left(O\right)$ and
for $\mathbf{H}\left(\mathbf{B}\right)$ are equivalent using the isomorphism
of eqs. \ref{eq:isom-cauchy-data-local} and \ref{eq:isom-cauchy-data-local2}%
. The first proof of outer regularity for $H\left(O\right)$ is in \cite%
{AraFF2} for the massive scalar free field and a class of regions more
general than double cones. A simple proof for massive or massless free
fields is given in \cite{LeyRobTesDforQFF} for the following class of
regions:

\begin{defn}
\label{def:strongly-contractible-region}A set $\mathbf{B}\subset\R%
^{d}$ is said strongly contractible around $0$ iff for any $\lambda$, $%
0<\lambda<1$, there exists an open neighborhood $\mathbf{A\subset}\R%
^{d}$ of $0$ such that $\lambda\left(\mathbf{B}+\mathbf{A}\right)\subset%
\mathbf{B}$. $\mathbf{B}$ is said strongly contractible if some of its
translated is strongly contractible around $0$.
\end{defn}

The proof in \cite{LeyRobTesDforQFF} uses dilation operators: for any $%
\lambda\in \R_{+}$, the dilation operator $D_{\lambda}f\left(x%
\right):= \lambda^{\frac{d-1}{2}} f\left(\lambda x\right)$ can be
extended to a bounded operator on $H $ (unitary iff the field is
massless) and $D_{\lambda}$ converges strongly
to $\mathbf{1}$. Therefore, $f\in \bigcap_{\mathbf{A}}\mathbf{H}\left(%
\mathbf{B+A}\right)$ implies that $D_{\lambda}f\in \mathbf{H}\left(\mathbf{B}%
\right)$ for any $\lambda <1$ and $f=\lim_{\lambda\rightarrow1_{-}}D_{\lambda}f\in \mathbf{H}%
\left(\mathbf{B}\right)$.

For the non-contractible regions used in Subsec. \ref{sub:massless-case} we
generalize the idea of \cite{LeyRobTesDforQFF} using more general
diffeomorphisms and prove here the following:

\begin{prop}
\label{pro:outer-regularity-O1-Oc}Let $O=C\left(\mathbf{B}\right)$ and $%
O_{1}=C\left(\mathbf{B}_{1}\right)$ be two double cones with basis $\mathbf{B%
}$ and $\mathbf{B}_{1}$ on the time-$0$ Cauchy surface, such that $\overline{%
O}\subset O_{1}$ ($\overline{\mathbf{B}}\subset\mathbf{B}_{1}$); the local
space $H\left(O_{1}\cap O^{\prime}\right)\simeq\mathbf{H}\left(\mathbf{%
B_{1}\backslash\overline{B}}\right)$ for the scalar free field (Def. \ref%
{def:1-p-Hilbert-space}) satisfies outer regularity.
\end{prop}

We want to prove that $H\left(O_{1}\cap
O^{\prime}\right)=\bigcap_{A}H\left(\left(O_{1}\cap
O^{\prime}\right)+A\right)$, where $A$ ranges in the set of open
neighborhoods of $0$ in $\R^{d+1}$, which is equivalent to
\[
\mathbf{H%
}\left(\mathbf{B_{1}\backslash\overline{B}}\right)=\bigcap_{\mathbf{A}}%
\mathbf{H}\left(\mathbf{\left(B_{1}\backslash\overline{B}\right)+A}\right)\,,
\]
where $\mathbf{A}$ ranges in the set of open neighborhoods of $0$ in $%
\R^{d}$. As dilations cannot map $\left(\mathbf{B}_{1}\backslash%
\overline{\mathbf{B}}\right)+\mathbf{A}$ in $\mathbf{B}_{1}\backslash%
\overline{\mathbf{B}},$ we need more general diffeomorphisms. For $\lambda$
in the neighborhood of $1$, let $\varphi_{\lambda}$ be a family of
diffeomorphisms which coincide with the identity outside a fixed compact
set. There are well defined constants $a_{\lambda}:=\sup_{\mathbf{x}%
}\left|\varphi_{\lambda}\left(\mathbf{x}\right)-\mathbf{x}\right|$
and $b_{\lambda} := \sup_{\mathbf{x}} \left|
\frac{\partial\varphi_{\lambda}}{\partial\mathbf{x}}
\left(\mathbf{x}\right)-\idop \right|$ (to simplify the notation
$\left|\cdot\right|$ indicates the modulus of numbers, the
Euclidean norm of vectors or the operator norm of matrices).
Suppose that $\varphi_{\lambda}$ converges to the identity as $%
\lambda\rightarrow1$, in the sense that
$a_{\lambda},b_{\lambda}\rightarrow0$.

\begin{prop}
Let $D_{\lambda}$ be the operator on $f_{0}\oplus f_{1}\in
C_{c}^{\infty}\left(\R^{d}\right)\tpr
\subset\mathbf{H}\left(\R^{d}\right)$ defined
by $D_{\lambda}f_{0,1}\left(x\right):=f_{0,1}\left(\varphi_{\lambda}\left(x%
\right)\right)$, then $D_{\lambda}$ can be extended to a bounded operator on
$\mathbf{H}\left(\R^{d}\right)$. Moreover, $%
D_{\lambda}\rightarrow\mathbf{1}$ strongly for $\lambda\rightarrow1$.
\end{prop}

\begin{proof}
Following Lemma 2.6.1. in \cite{HorLPDO}, for $s<1$ there is a constant $%
A_{s}$ such that, for any $f\in
C_{c}^{\infty}\left(\R^{d}\right)$,
\begin{multline*}
\int\frac{\left|f\left(\mathbf{x}\right)-f\left(\mathbf{y}\right)\right|^{2}%
}{\left|\mathbf{x}-\mathbf{y}\right|^{d+2s}}d\mathbf{x}d\mathbf{y}=\int\frac{%
\left|f\left(\mathbf{z}+\mathbf{y}\right)-f\left(\mathbf{y}\right)\right|^{2}%
}{\left|\mathbf{z}\right|^{d+2s}}d\mathbf{z}d\mathbf{y}= \\
=\int\frac{\left|\left(e^{i\mathbf{pz}}-1\right)\widehat{f}\left(\mathbf{p}%
\right)\right|^{2}}{\left|\mathbf{z}\right|^{d+2s}}d\mathbf{z}d\mathbf{p}%
=A_{s}\int\left|\widehat{f}\left(\mathbf{p}\right)\right|^{2}\left|\mathbf{p}%
\right|^{2s}d\mathbf{p}
\end{multline*}
because for any $\mathbf{z}$ the Fourier transform of the function $\mathbf{y%
}\mapsto f\left(\mathbf{z}+\mathbf{y}\right)-f\left(\mathbf{y}\right)$ is $%
\mathbf{p}\mapsto\left(e^{i\mathbf{pz}}-1\right)\widehat{f}\left(\mathbf{p}%
\right)$ and $\int\frac{\left|\left(e^{i\mathbf{pz}}-1\right)\right|^{2}}{%
\left|\mathbf{z}\right|^{d+2s}}d\mathbf{z}$ is an homogeneous function of $%
\mathbf{p}$ of degree $2s$, thus equal to
$A_{s}\left|\mathbf{p}\right|^{2s}$.
Therefore, as $\left\Vert f\right\Vert _{0,1}^{2}=\int\left|\widehat{f}%
\left(\mathbf{p}\right)\right|^{2}\left|\mathbf{p}\right|^{\pm1}d\mathbf{p}$%
, with $\mathbf{x}_{\lambda}^{\prime}:=\varphi_{\lambda}\left(\mathbf{x}%
\right)$ we have
\begin{multline*}
\left\Vert f\circ\varphi_{\lambda}\right\Vert _{0,1}^{2} =
A_{\pm}^{-1}\int \frac{\left|
f\left(\mathbf{x}_{\lambda}^{\prime}\right)
-f\left(\mathbf{y}_{\lambda}^{\prime} \right)\right| ^{2}}
{\left|\mathbf{x}-\mathbf{y}\right|^{d\pm1}} d\mathbf{x}d\mathbf{y}\leq \\
\leq A_{\pm}^{-1} {\left( 1-b_{\lambda} \right)^{-2d}}
\left(1+b_{\lambda}\right)^{d\pm1}\int \frac{\left|
f\left(\mathbf{x}_{\lambda}^{\prime}\right) -
f\left(\mathbf{y}_{\lambda}^{\prime}\right)\right| ^{2}}
{\left|\mathbf{x}_{\lambda}^{\prime}-\mathbf{\mathbf{y}
_{\lambda}^{\prime}}\right|^{d\pm1}}
d\mathbf{x}_{\lambda}^{\prime}d\mathbf{y_{\lambda}^{\prime}} =
 const.\left\Vert f\right\Vert _{0,1}^{2}
\end{multline*}
where we used the estimate $\sup_{\mathbf{x}}\left|
\mathop{det}\left(
\frac{\partial\mathbf{x}}{\partial\mathbf{x}_{\lambda}^{\prime}}
\right)\right| \leq \sup_{\mathbf{x}} \left|
\frac{\partial\mathbf{x}}{\partial\mathbf{x}_{\lambda}^{\prime}}
\right|^d \leq \left(1-b_{\lambda}\right) ^{-d}$  (from
$\frac{\partial\mathbf{x}_{\lambda}^{\prime}}{\partial\mathbf{x}}=\idop
+ \left(
\frac{\partial\mathbf{x}_{\lambda}^{\prime}}{\partial\mathbf{x}} -
\idop \right)$ with $\left|
\frac{\partial\mathbf{x}_{\lambda}^{\prime}}{\partial\mathbf{x}} -
\idop \right| \leq b_{\lambda} <1$, follows the bound for the
inverse $\left|
\frac{\partial\mathbf{x}}{\partial\mathbf{x}_{\lambda}^{\prime}}
\right| \leq \frac{1}{1-b_{\lambda}}$) and
$\sup_{\mathbf{x},\mathbf{y%
}}\frac{\left|\mathbf{x}_{\lambda}^{\prime}-\mathbf{\mathbf{y}
_{\lambda}^{\prime}}\right|}{\left|\mathbf{x}-\mathbf{y}\right|}
\leq \sup_{\mathbf{x}} \left|
\frac{\partial\mathbf{x}_{\lambda}^{\prime}}{\partial\mathbf{x}}
\left(\mathbf{x}\right) \right| \leq 1+b_{\lambda}$. This proves
that the operators $D_{\lambda}$ are bounded, with
$\left\Vert D_{\lambda}\right\Vert ^{2}\leq %
{\left( 1-b_{\lambda} \right)^{-2d}}
\left(1+b_{\lambda}\right)^{d\pm1}$.

For a fixed $f$, defining
$f_{\lambda}:=f\circ\varphi_{\lambda}-f$, $\supp f_{\lambda}$ is
contained in a fixed compact set for any $\lambda$ and $\left\Vert
f_{\lambda}\right\Vert _{0,1}^{2}=A_{\pm}^{-1} \int \frac{\left|
f_{\lambda}\left(\mathbf{x}\right) -
f_{\lambda}\left(\mathbf{y}\right)\right|^{2}}
{\left|\mathbf{x}-\mathbf{y}\right|^{d\pm1}}
d\mathbf{x}d\mathbf{y} \leq const.\sup_{\mathbf{x}}\left|
\frac{\partial{f}_\lambda}{\partial\mathbf{x}}\left(\mathbf{x}\right)
\right|^{2} \rightarrow 0$ because
\begin{multline*}
\sup_{\mathbf{x}} \left|
\frac{\partial{f}_{\lambda}}{\partial\mathbf{x}}\left(\mathbf{x}
\right)\right| =
 \sup_{\mathbf{x}} \left|
\left(\frac{\partial{f}}{\partial\mathbf{x}}\left(
\mathbf{x}_{\lambda}^{\prime}\left(\mathbf{x}\right)\right) -%
\frac{\partial{f}}{\partial\mathbf{x}}\left(\mathbf{x}\right)\right)%
\frac{\partial\mathbf{x}_{\lambda}^{\prime}}{\partial\mathbf{x}}
\left(\mathbf{x}\right) + \frac{\partial{f}}{\partial\mathbf{x}}%
\left(\mathbf{x}\right)
\left(\frac{\partial\mathbf{x}_{\lambda}^{\prime}}{\partial\mathbf{x}}\left(\mathbf{x}
\right)-\idop \right)\right|%
\leq \\
\leq \sup_{\mathbf{x}} \left\{%
\sup_{\mathbf{y}} \left|
\frac{\partial^2{f}}{\partial\mathbf{x}^2}\left(\mathbf{y}\right) \right|%
\left|\mathbf{x}_{\lambda}^{\prime}-\mathbf{x} \right|%
\left|
\frac{\partial\mathbf{x}_{\lambda}^{\prime}}{\partial\mathbf{x}}
\left(\mathbf{x}\right) \right|%
+ \left| \frac{\partial{f}}
{\partial\mathbf{x}} \left(\mathbf{x}\right) \right|%
\left| \frac{\partial\mathbf{x}_{\lambda}^{\prime}}
{\partial\mathbf{x}}\left(\mathbf{x} \right)-\idop \right|
\right\} \leq\\
\leq \sup_{\mathbf{x}} \left|
\frac{\partial^2{f}}{\partial\mathbf{x}^2}\left(\mathbf{x}\right) \right|%
a_{\lambda} \left(1+b_{\lambda}\right) %
+ \sup_{\mathbf{y}} \left| \frac{\partial{f}}{\partial\mathbf{x}}
\left(\mathbf{y}\right) \right| b_{\lambda} \rightarrow 0
\end{multline*}
This proves that $\left\Vert
D_{\lambda}f-f\right\Vert \rightarrow0$ for $f\in C_{c}^{\infty}\left(%
\R^{d}\right)\otimes\R^{2}$ and, by the above uniform bound
on $\left\Vert D_{\lambda}\right\Vert $, for any $f\in \mathbf{H}\left(%
\R^{d}\right)$.
\end{proof}

It is clear that such a family of diffeomorphisms $\varphi_{\lambda}$ can be
chosen so that for any $\lambda>1$ there is an open set $\mathbf{A}$ such
that $\varphi_{\lambda}\left(\left(\mathbf{B}_{1}\backslash\overline{\mathbf{%
B}}\right)+\mathbf{A}\right)\subset\mathbf{B}_{1}\backslash\overline{\mathbf{%
B}}$. As a consequence, $f\in \bigcap_{\mathbf{A}}\mathbf{H}\left(\mathbf{%
\left(B_{1}\backslash\overline{B}\right)+A}\right)$ implies that, for any $%
\lambda>1$, $\varphi_{\lambda}\left(f\right)\in \mathbf{H}\left(\mathbf{%
B_{1}\backslash\overline{B}}\right)$ and, taking the limit $%
\lambda\rightarrow1$, $f\in \mathbf{H}\left(\mathbf{B_{1}\backslash\overline{%
B}}\right)$.

\section{\label{app:Differential-forms}On Differential Forms and the
Propagator}

Let $M$ be a globally hyperbolic $d+1$-dimensional manifold and $\mathbf{%
\cs}$ any Cauchy (smooth) surface. The Lorentzian metric on $M$
and the induced Riemannian one on $\cs$ define canonically a
volume form $\omega$ and an inner product on the external algebra
of the cotangent space: $a_{p}\cdot b_{p}\in \R$ with
$a_{p},b_{p}\in \Lambda^{\ast}\left(T_{p}^{\ast}M\right)$ and
$\mathbf{a}_{p}\cdot\mathbf{b}_{p} \in \R$ with $\mathbf{a}_{p},\mathbf{b%
}_{p}\in \Lambda^{\ast}\left(T_{p}^{\ast}\cs\right)$. The inner
product on forms with compact support $\Omega_{c}^{k}\left(M\right)$ and $%
\Omega_{c}^{k}\left(\cs\right)$ is the integral of the
point-wise inner product $a\cdot b\in \Omega_{c}^{0}\left(M\right)$: $%
\left\langle a,b\right\rangle :=-\int\left(a\cdot b\right)\omega$. It is
indefinite on the first space and positive definite on the second (the minus
sign is necessary according to the Lorentzian signature $\left(+-...-\right)$
of the metric). The Hodge star operation on differential forms $%
\ast:\Omega_{}^{k}\left(M\right)\rightarrow\Omega_{}^{d+1-k}\left(M\right)$,
$\ast:\Omega_{}^{k}\left(\cs\right)\rightarrow\Omega_{}^{d-k}%
\left(\cs\right)$, is defined point-wise according to $%
\left(a_{p}\cdot b_{p}\right)\omega=a_{p}\wedge\ast b_{p}$, then $%
\left\langle a,b\right\rangle =-\int a\wedge\ast b$. On $\Omega_{}^{k}%
\left(M\right)$, $\ast^{2}=-\left(-1\right)^{kd}$ (the extra minus is due to
the above signature of the metric) and on $\Omega_{}^{k}\left(\cs%
\right)$, $\ast^{2}=\left(-1\right)^{k\left(d+1\right)}$.

Through this inner product, $\Omega_{c}^{k}\left(M\right)$ is a
subset of the space of $k$-currents (set of distributions valued
$k$-forms \cite{SchwartzTdD} that we denote $\Omega_{}^{\prime
k}\left(M\right)$: it is the space of continuous linear functional
on $\Omega_{c}^{k}\left(M\right)$ with respect to the standard
topology as a test forms' space). We indicate again with the same
symbol $\left\langle a,b\right\rangle $ also the action of the
current $a$ on the form $b$.

From the relation $\int a\wedge\ast
b=\left(-1\right)^{k\left(d+1-k\right)}\int\ast a\wedge b$, for
$a,b\in \Omega_{c}^{k}\left(M\right)$, one deduces that the
transpose of the
exterior differentiation $d:\Omega_{}^{k}\left(M\right)\rightarrow%
\Omega_{}^{k+1}\left(M\right)$ is $\delta:=\left(-1\right)^{\left(d+1%
\right)k}\ast d\ast:\Omega_{}^{\prime
k+1}\left(M\right)\rightarrow\Omega_{}^{\prime k}\left(M\right)$.
Similarly $\delta:=\left(-1\right)^{dk}\ast d\ast$ on $\cs$. For
simplicity, we use the same letters $d$, $\delta$ and $\ast$ for
operators on the different spaces $\Omega_{}^{k}\left(M\right)$,
$\Omega_{}^{k}\left(\cs\right)$, $\Omega_{}^{\prime
k}\left(M\right)$, $\Omega_{}^{\prime k}\left(\cs\right)$. The
closed elements with respect to $d$
and $\delta$ (cocycles and cycles) are indicated respectively by $%
Z^{d}\Omega_{}^{k}\left(M\right)$ and $Z^{\delta}\Omega_{}^{k}\left(M\right)$%
, the (co)boundary elements respectively by $B^{d}\Omega_{}^{k}\left(M\right)
$ and $B^{\delta}\Omega_{}^{k}\left(M\right)$.

Let $i_{X}:\Omega_{}^{k}\left(M\right)\rightarrow\Omega_{}^{k-1}\left(M%
\right)$ be the contraction of a differential form with a vector
field $X$; through the metric, we can identify a vector field $X$
with a $1$-form $X^{\ast}$ and vice versa, such that $\left\langle
X^{\ast}\wedge a,b\right\rangle =\left\langle
a,i_{X}b\right\rangle $ for any couple of forms $a$ and $b$, or such that $%
\left\langle X^{\ast},a\right\rangle =a\left(X\right)$ for any $a\in
\Omega_{}^{1}\left(M\right)$. The Lie derivative along a vector field $X$ is
$L_{X}=i_{X}d+di_{X}$.

Given a Cauchy surface $\cs$ let $j:\cs%
\hookrightarrow M$ be the injection and $j^{\ast}:\Omega_{}^{k}\left(M%
\right)\rightarrow\Omega_{}^{k}\left(\cs\right)$ the pull-back of
forms. Clearly, $j^{\ast}$ is the restriction to $\cs$ of the
tangent component of the differential form. The restriction to $\mathbf{%
\cs}$ of the normal component of the differential form is $%
j^{\ast}i_{X_{n}}$, where $X_{n}$ is any normalized vector field
orthogonal to $\cs$, and it coincides
with$\left(-1\right)^{d-k}\ast
j^{\ast}\ast:\Omega_{}^{k}\left(M\right)\rightarrow\Omega_{}^{k-1}\left(%
\cs\right)$; it can be verified in a point $p\in \cs$ using local
coordinates $x^{i}$, which can be chosen to be orthonormal in $p$
and such that $x^{0}=0$ on $\cs$.

We define two operators $\tilde{\rho}_{0},\tilde{\rho}_{1}:\Omega_{}^{k}%
\left(M\right)\rightarrow\Omega_{}^{k}\left(\cs\right)\oplus
\Omega_{}^{k-1}\left(\cs\right)$, extending those already
defined on $0$-forms (eq. \ref{eq:rho_0-rho_1}): on $\Omega_{}^{k}\left(M%
\right)$%
\begin{equation*}
\tilde{\rho}_{0}:=j^{\ast}\oplus \left(-1\right)^{d-k}\ast
j^{\ast}\ast,\,\,\,\,\tilde{\rho}_{1}:=\left(-1\right)^{d+1-k}\left(\left[%
\ast j^{\ast}\ast,d\right]\oplus \left[j^{\ast},\delta\right]\right).
\end{equation*}
They have two components: the restriction to $\cs$ (of the normal
derivative) of the tangent and of the normal component of the
differential form (the normal derivative on $\cs$ can be written
as $\left(-1\right)^{d+1-k}\left[\ast j^{\ast}\ast,d\right]
=j^{\ast}L_{X_{n}}$). We define
also simpler operators that are used in eq. \ref{eq:EM-E-cauchy-data2}%
, $\rho_{0},\rho_{1}:\Omega_{}^{k}\left(M\right)\rightarrow\Omega_{}^{k}%
\left(\cs\right)$,
\begin{equation}
\rho_{0}:=j^{\ast},\,\,\,\rho_{1}:=\left(-1\right)^{d+1-k}\ast j^{\ast}\ast d
\label{eq:EM-rho_0-rho_1}
\end{equation}
and, as in the scalar case, their transpose on the space of currents.

The propagator $E$ and the operator $P$, already defined for the scalar case
(see eq. \ref{eq:E-def} and \ref{eq:P-def}) as convolutions with solutions
of the wave equation, can be extended to differential forms with compact
support.

\begin{prop}
The following equality (see \ref{eq:E-cauchy-data}) holds on $%
\Omega_{c}^{k}\left(M\right)$:
\begin{eqnarray}
E & = & E\left(\tilde{\rho}_{0}^{\prime}\tilde{\rho}_{1}-\tilde{\rho}%
_{1}^{\prime}\tilde{\rho}_{0}\right)E=  \label{eq:EM-E-cauchy-data1} \\
& = &
E\left(\rho_{0}^{\prime}\rho_{1}-\rho_{1}^{\prime}\rho_{0}\right)E+\ast
E\left(\rho_{0}^{\prime}\rho_{1}-\rho_{1}^{\prime}\rho_{0}\right)E\ast.
\label{eq:EM-E-cauchy-data2}
\end{eqnarray}
\end{prop}

\begin{proof}
Equation \ref{eq:EM-E-cauchy-data1} is true on
$\Omega_{}^{0}\left(M\right)$ where it coincides with
\ref{eq:E-cauchy-data}; it can be proven by induction on the
degree of the differential forms showing that both sides of eq.
\ref{eq:EM-E-cauchy-data1} commute with $i_{X}$, where $X$ is any
constant vector field on $M\subset\R^{d+1}$. As $E$ commutes with $%
i_{X}$, one has to check that also
$\tilde{\rho}_{0}^{\prime}\tilde{\rho}_{1} $ and
$\tilde{\rho}_{1}^{\prime}\tilde{\rho}_{0}$ does. One can compute,
using local coordinates in the neighborhood of any $p\in \cs$,
that $\left(j^{\ast}a\right)_{p}\cdot\left(j^{\ast}b\right)_{p}+\left(j^{%
\ast}\ast a\right)_{p}\cdot\left(j^{\ast}\ast
b\right)_{p}=j^{\ast}\left(a\cdot b\right)_{p}$ and thus

\begin{equation*}
\left\langle j^{\ast}a,j^{\ast}b\right\rangle _{\cs%
}+\left\langle j^{\ast}\ast a,j^{\ast}\ast b\right\rangle _{\cs%
}=\int_{\cs}\ast j^{\ast}\left(a\cdot b\right)
\end{equation*}
Let us indicate with $L_{n}:=L_{X_{n}}$ the Lie derivative along
any normalized vector field $X_{n}$ orthogonal to $\cs$. $L_{n}$
commutes with $X$ if it is a constant vector field and one can compute%
\begin{multline*}
\left\langle a,\tilde{\rho}_{0}^{\prime}\tilde{\rho}_{1}i_{X}b\right\rangle
=\left\langle j^{\ast}a,j^{\ast}L_{n}i_{X}b\right\rangle _{\cs%
}+\left\langle j^{\ast}\ast a,j^{\ast}L_{n}\ast i_{X}b\right\rangle _{%
\cs}= \\
=\left\langle j^{\ast}a,j^{\ast}i_{X}L_{n}b\right\rangle _{\cs%
}+\left\langle j^{\ast}\ast a,j^{\ast}\ast
i_{X}L_{n}b\right\rangle _{\cs} = \\
=\int_{\cs}\ast j^{\ast}\left(a\cdot i_{X}L_{n}b\right)=
\int_{\cs}\ast j^{\ast}\left(\left(X^{\ast}\wedge
a\right)\cdot\left(L_{n}b\right)\right) = \\
=\left\langle
j^{\ast}\left(X^{\ast}\wedge a\right),j^{\ast}L_{n}b\right\rangle _{\mathbf{%
\cs}}+\left\langle j^{\ast}\ast \left(X^{\ast}\wedge
a\right),j^{\ast}\ast L_{n}b\right\rangle _{\cs}= \\
=\left\langle \tilde{\rho}_{0}\left(X^{\ast}\wedge a\right),\tilde{\rho}%
_{1}b\right\rangle _{\cs}=\left\langle X^{\ast}\wedge a,\tilde{%
\rho}_{0}^{\prime}\tilde{\rho}_{1}b\right\rangle =\left\langle a,i_{X}\tilde{%
\rho}_{0}^{\prime}\tilde{\rho}_{1}b\right\rangle .
\end{multline*}
Similarly one checks that $\left\langle a,\tilde{\rho}_{1}^{\prime}\tilde{%
\rho}_{0}i_{X}b\right\rangle =\left\langle a,i_{X}\tilde{\rho}_{1}^{\prime}%
\tilde{\rho}_{0}b\right\rangle $.

To prove the second equality \ref{eq:EM-E-cauchy-data2}, we compute%
\begin{multline*}
\left\langle \tilde{\rho}_{0}a,\tilde{\rho}_{1}b\right\rangle _{\mathbf{%
\cs}}=\left\langle j^{\ast}a,\left[\ast j^{\ast}\ast,d\right]%
b\right\rangle _{\cs}-\left\langle \ast j^{\ast}\ast a,\left[%
j^{\ast},\delta\right]b\right\rangle _{\cs}= \\
=\left\langle j^{\ast}a,\ast j^{\ast}\ast db\right\rangle _{\cs%
}-\left\langle j^{\ast}a,d\ast j^{\ast}\ast b\right\rangle _{\cs%
}-\left\langle \ast j^{\ast}\ast a,j^{\ast}\delta b\right\rangle _{\mathbf{%
\cs}}+\left\langle \ast j^{\ast}\ast a,\delta j^{\ast}b\right\rangle _{%
\cs}= \\
=\left\langle j^{\ast}a,\ast j^{\ast}\ast db\right\rangle _{\cs%
}-\left\langle \ast j^{\ast}\ast a,j^{\ast}\delta b\right\rangle _{\mathbf{%
\cs}}-\left\langle j^{\ast}a,d\ast j^{\ast}\ast b\right\rangle _{\mathbf{%
\cs}}+\left\langle d\ast j^{\ast}\ast a,j^{\ast}b\right\rangle _{\mathbf{%
\cs}}
\end{multline*}
therefore, taking the antisymmetric part in the exchange $0\leftrightarrow1$,%
\begin{multline*}
\left\langle \tilde{\rho}_{0}a,\tilde{\rho}_{1}b\right\rangle _{\mathbf{%
\cs}}-\left\langle \tilde{\rho}_{0}b,\tilde{\rho}_{1}a\right\rangle _{%
\cs}= \\
=\left\langle j^{\ast}a,\ast j^{\ast}\ast db\right\rangle _{\cs%
}-\left\langle j^{\ast}b,\ast j^{\ast}\ast da\right\rangle _{\cs%
}+\left\langle \ast j^{\ast}\ast b,j^{\ast}\delta a\right\rangle _{\mathbf{%
\cs}}-\left\langle \ast j^{\ast}\ast a,j^{\ast}\delta b\right\rangle _{%
\cs}= \\
=\left\langle \rho_{0}a,\rho_{1}b\right\rangle _{\cs%
}-\left\langle \rho_{0}b,\rho_{1}a\right\rangle _{\cs%
}+\left\langle \rho_{0}\ast a,\rho_{1}\ast b\right\rangle _{\cs%
}-\left\langle \rho_{0}\ast b,\rho_{1}\ast a\right\rangle _{\cs}
\end{multline*}
and finally $\tilde{\rho}_{0}^{\prime}\tilde{\rho}_{1}-\tilde{\rho}%
_{1}^{\prime}\tilde{\rho}_{0}=\rho_{0}^{\prime}\rho_{1}-\rho_{1}^{\prime}%
\rho_{0}+\ast\left(\rho_{0}^{\prime}\rho_{1}-\rho_{1}^{\prime}\rho_{0}%
\right)\ast$.
\end{proof}

\



\begin{thebibliography}{10}

\bibitem{AraFF1}
H.~Araki.
\newblock A lattice of von {N}eumann algebras associated with the quantum
  theory of a free {B}ose field.
\newblock {\em J. Math. Phys.}, 4:1343--1362, 1963.

\bibitem{AraFF2}
H.~Araki.
\newblock von {N}eumann algebras of local observables for free scalar field.
\newblock {\em J. Math. Phys.}, 5:1--13, 1964.

\bibitem{AraYamQEofQFS}
H.~Araki and S.~Yamagami.
\newblock On quasi-equivalence of quasifree states of the canonical commutation
  relations.
\newblock {\em Publ. RIMS, Kyoto Univ.}, 18:283--338, 1982.

\bibitem{BenNicLAforSFF&EMFF}
G.~Benfatto and F.~Nicol{\`o}.
\newblock The local von {N}eumann algebras for the massless scalar free field
  and the free electromagnetic field.
\newblock {\em J. Mathematical Phys.}, 19(3):653--660, 1978.

\bibitem{SanBerEoCStSCS}
A.~N. Bernal and M.~S\'anchez.
\newblock A note on the extendability of compact hypersurfaces to smooth
  {C}auchy hypersurfaces.
\newblock {\em arXiv:gr-qc/0507018}, 2005.

\bibitem{OAQSM2}
O.~Bratteli and D.~W. Robinson.
\newblock {\em Operator Algebras and Quantum Statistical Mechanics. {I}{I}}.
\newblock Springer-Verlag, Berlin-Heidelberg-New York, 1979.

\bibitem{BucLQFwNTI}
D.~Buchholz.
\newblock On the structure of local quantum fields with nontrivial interaction.
\newblock In {\em Proceedings of the International Conference on Operator
  Algebras, Ideals, and their Applications in Theoretical Physics (Leipzig,
  1977)}, pages 146--153, Leipzig, 1978. Teubner.

\bibitem{BucSofQED}
D.~Buchholz.
\newblock The physical state space of quantum electrodynamics.
\newblock {\em Comm. Math. Phys.}, 85(1):49--71, 1982.

\bibitem{DelSofAofFS}
G.~Dell'Antonio.
\newblock Structure of the algebras of some free systems.
\newblock {\em Commun. Math. Phys.}, 9:81--117, 1968.

\bibitem{DimLO}
J.~Dimock.
\newblock Algebras of local observables on a manifold.
\newblock {\em Commun. Math. Phys.}, 77:219--228, 1980.

\bibitem{DimEMF}
J.~Dimock.
\newblock Quantized electromagnetic field on a manifold.
\newblock {\em Rev. Math. Phys.}, 4(2):223--233, 1992.

\bibitem{DHR1}
S.~Doplicher, R.~Haag, and J.~E. Roberts.
\newblock Fields, observables and gauge transformations {I}.
\newblock {\em Commun. Math. Phys.}, 13:1--23, 1969.

\bibitem{DHR2}
S.~Doplicher, R.~Haag, and J.~E. Roberts.
\newblock Fields, observables and gauge transformations {I}{I}.
\newblock {\em Commun. Math. Phys.}, 15:173--200, 1969.

\bibitem{DHR3}
S.~Doplicher, R.~Haag, and J.~E. Roberts.
\newblock Local observables and particle statistics {I}.
\newblock {\em Commun. Math. Phys.}, 23:199--230, 1971.

\bibitem{DHR4}
S.~Doplicher, R.~Haag, and J.~E. Roberts.
\newblock Local observables and particle statistics {I}{I}.
\newblock {\em Commun. Math. Phys.}, 35:49--85, 1974.

\bibitem{EckOstDforFBF}
J.-P. Eckmann and K.~Osterwalder.
\newblock An application of {T}omita's theory of modular {H}ilbert algebras:
  Duality for free {B}ose fields.
\newblock {\em J. Funct. Anal.}, 13:1--12, 1973.

\bibitem{LQP}
R.~Haag.
\newblock {\em Local Quantum Physics}.
\newblock Texts and Monographs in Physics. Springer-Verlag, Berlin-Heidelberg,
  1992.

\bibitem{HisDforLAinFQFT}
P.~D. Hislop.
\newblock A simple proof of duality for local algebras in free quantum field
  theory.
\newblock {\em J. Math. Phys.}, 27(10):2542--2550, 1986.

\bibitem{MR89j:81088}
P.~D. Hislop.
\newblock Conformal covariance, modular structure, and duality for local
  algebras in free massless quantum field theories.
\newblock {\em Ann. Physics}, 185(2):193--230, 1988.

\bibitem{HisLonMSofLAofFMSFT}
P.~D. Hislop and R.~Longo.
\newblock Modular structure of the local algebras associated with the free
  massless scalar field theory.
\newblock {\em Commun. Math. Phys.}, 84(1):71--85, 1982.

\bibitem{HorLPDO}
L.~H{\"o}rmander.
\newblock {\em Linear partial differential operators}.
\newblock Third revised printing. Die Grundlehren der mathematischen
  Wissenschaften, Band 116. Springer-Verlag New York Inc., New York, 1969.

\bibitem{LeyRobTesDforQFF}
P.~Leyland, J.~E. Roberts, and D.~Testard.
\newblock Duality for quantum free fields.
\newblock unpublished, 1978.

\bibitem{LudRobLQE&AVS}
C.~Lueders and J.~E. Roberts.
\newblock Local quasiequivalence and adiabatic vacuum states.
\newblock {\em Commun. Math. Phys.}, 134:29--63, 1990.

\bibitem{ManCCR}
J.~Manuceau.
\newblock C*-alg\`{e}bre de relations de commutation.
\newblock {\em Ann. Inst. H. Poincar{\'e}}, VIII:139--161, 1968.

\bibitem{OstDforFBF}
K.~Osterwalder.
\newblock Duality for free {B}ose fields.
\newblock {\em Commun. Math. Phys.}, 29:1--14, 1973.

\bibitem{SW-TSiQFT}
P.~Sadowski and S.~L. Woronowicz.
\newblock Total sets in quantum field theory.
\newblock {\em Rep. Mathematical Phys.}, 2(2):113--120, 1971.

\bibitem{SchwartzTdD}
L.~Schwartz.
\newblock {\em Th\`eorie des Distributions}.
\newblock Paris, Hermann, 1966.

\end{thebibliography}

\end{document}